%% file: arxiv.tex
\documentclass[12pt]{article}

\usepackage{graphicx, subfigure}
\usepackage[top=1in,bottom=1in,left=1in,right=1in]{geometry}

\usepackage{amsfonts, amsthm, amsmath, amssymb, bbm}

\usepackage{times}
\usepackage{bm}
\usepackage{natbib}
\usepackage{moreverb,url}
\usepackage{multirow}
\usepackage{float}
\usepackage{mathtools}
\mathtoolsset{showonlyrefs}
\usepackage{color}
\definecolor{darkred}{RGB}{100,0,0}
\definecolor{darkgreen}{RGB}{0,100,0}
\definecolor{darkblue}{RGB}{0,0,150}
\usepackage{prodint}

\usepackage{hyperref}
\hypersetup{colorlinks=true, linkcolor=darkred, citecolor=darkgreen, urlcolor=darkblue}

\usepackage[plain,noend]{algorithm2e}

\input{notation.tex}

\title{\bf{Proximal Survival Analysis to Handle Dependent Right Censoring}}

\author{Andrew Ying}

\date{}

\begin{document}

\maketitle

\begin{abstract}
Many epidemiological and clinical studies aim at analyzing a time-to-event endpoint. A common complication is right censoring. In some cases, it arises because subjects are still surviving after the study terminates or move out of the study area, in which case right censoring is typically treated as independent or non-informative. Such an assumption can be further relaxed to conditional independent censoring by leveraging possibly time-varying covariate information, if available, assuming censoring and failure time are independent among covariate strata. In yet other instances, events may be censored by other competing events like death and are associated with censoring possibly through prognoses. Realistically, measured covariates can rarely capture all such associations with certainty. For such dependent censoring, often covariate measurements are at best proxies of underlying prognoses. In this paper, we establish a nonparametric identification framework by formally admitting that conditional independent censoring may fail in practice and accounting for covariate measurements as imperfect proxies of underlying association. The framework suggests adaptive estimators which we give generic assumptions under which they are consistent, asymptotically normal, and doubly robust. We illustrate our framework with concrete settings, where we examine the finite-sample performance of our proposed estimators via a Monte-Carlo simulation and apply them to the SEER-Medicare dataset.
\end{abstract}




\section{Introduction}
Many epidemiological and clinical studies aim at analyzing a time to an event. A central complication that arises though \citep{kalbfleisch2011statistical, andersen2012statistical}, is the right censoring when event of interest is censored when censoring happens before failure time. In some cases, right censoring arises because subjects are still surviving after the study terminates or they may move out of the study area and therefore cannot be followed. In yet other instances, individuals may be withdrawn because of worsening prognoses or death. As is intuitively apparent, censoring may introduce bias into the study without careful treatment. 

To handle right censoring, the ``independent censoring'' assumption, or known as the ``non-informative censoring'' assumption \citep{kalbfleisch2011statistical}, is typically adopted in the literature, which requires the censoring time to be completely independent of the failure time. Under this assumption, individuals who are censored can be represented by those who remain under observation. ``Independent censoring'' ought to hold when censoring is determined before the start of the study, known as administrative censoring. Under independent censoring, the Kaplan-Meier estimator \citep{kaplan1958nonparametric}, Nelson-Aalen estimator, and logrank test are typically used to conduct valid inference for survival analysis. 

``Independent censoring'' fails to hold when censoring is correlated to failure time possibly via some unknown censoring mechanism. In that case, an important generalization of ``independent censoring'' is the ``conditional independent censoring'' assumption, where one leverages the measured (possibly time-dependent) covariates and claims that they are sufficiently enriched to render censoring and failure time independent within covariates strata. This assumption was also formalized as ``coarsening at random'' in the coarsened data literature \citep{van2003unified, rotnitzky2005inverse, tsiatis2006semiparametric} when treating right censoring as monotone coarsening. Under this assumption, the proportional hazards model \citep{cox1972regression, cox1975partial} and the additive hazards model \citep{aalen1980model, aalen1989linear} are commonly adopted models. Alternatively, an inverse probability weighting estimator and augmented inverse probability weighting estimator \citep{van2003unified, rotnitzky2005inverse, tsiatis2006semiparametric} can also be used to overcome censoring. 

However, ``conditional independent censoring'' is empirically untestable and must be taken on faith even with substantial subject matter knowledge at hand. Practically, measured covariates can hardly capture the underlying censoring mechanism perfectly. Realistically one can only hope that covariate measurements are at best imperfect proxies of the true underlying censoring mechanism operating in a given time-to-event study. Such acknowledgment invalidates any survival claim made on the basis of the ``conditional independent censoring'' assumption. 
The validity of the ``conditional independent censoring'' assumption is especially subject to questioning for treating competing events, that is, when the failure of an individual may be one of several distinct failures, as censoring. This is because individuals who experience competing events (treated as right censored) typically have different prognoses for the primary event, compared to subjects who neither were censored nor experienced competing events. Such a censoring mechanism depending heavily on prognoses is apparently difficult to be fully captured in practice, which may not be totally explained away by measured covariates. Therefore treating competing events as ``conditional independent censoring'' is unreasonable. In such cases, only some specific alternative quantities on the primary event involving the censoring variable can be inferred, for example, subdistribution hazard \citep{fine1999proportional, ying2019two} or cause-specific hazard function, whose definitions depend on competing events. 

In this work, instead of relying on either ``independent censoring'', ``conditional independent censoring'' on basis of measured covariates, or considering any alternative quantities, we develop a proximal approach to nonparametrically identify any finite-dimensional parameter based on the event of interest with possibly time-varying covariates, by borrowing ideas from ``proximal causal inference'' \citep{miao2018identifying, deaner2018proxy, shi2020selective, tchetgen2020introduction, singh2020kernel, imbens2021controlling,  cui2020semiparametric, ying2021proximal, kallus2021causal, mastouri2021proximal, dukes2021proximal, qi2021proximal, ying2022proximal}, which was recently developed to nonparametrically identify causal effects. Assuming that there are some latent factors rendering censoring ``independent'', our proximal survival analysis framework essentially requires that the analyst has measured covariates, that can be classified into three bucket types: 1) variables that may be directly associated with both the censoring and failure time; 2) potentially censoring-inducing proxies; and 3) potentially event-inducing proxies. A proxy of type 2) is associated with the failure time only through the latent factors, but allowed to be directly associated with the censoring time; while a proxy of type 3) is associated with the censoring only through the latent factors, but allowed to be directly associated with the failure time.

We give examples to illustrate definitions of proxies. In a longitudinal study on Alzheimer's disease \citep{p2012honolulu}, the event of interest is moderate impairment, and the right censoring includes loss to follow-up and death. Typically, loss to follow-up is treated as independent censoring and thus existing methods can readily apply. Death, on the other hand, is usually considered a competing event and hence dependent. Fortunately, the available data include important covariates known to be associated with both moderate impairment and death, including demographic, clinical, laboratory, other medication use, and self-reported health status updated over time. Among measured covariates, one may claim that grip strength is only associated with time to moderate impairment through some latent factors like underlying health conditions and thus is a proxy of type 2). The measurement of APOE genotype is directly associated with moderate impairment but otherwise only associated with death through underlying health conditions as well, which can be categorized as a proxy of type 3). The rest covariates like alcohol intake, heart rate, and education can be classified into type 1). As another example of prostate cancer \citep{byar1980choice}, patients with late-stage prostate cancer were followed for 60 months. The event of interest is time to death due to prostate cancer, and the right censoring includes alive and death due to other causes (mainly cardiovascular diseases). Baseline covariates were collected, including age, weight index, performance rating, history of cardiovascular diseases, serum hemoglobin, size of the primary tumor, systolic blood pressure, and diastolic blood pressure. Among these, one may say that history of cardiovascular diseases, systolic blood pressure, and diastolic blood pressure are proxies of type 2), whereas serum hemoglobin and size of the primary tumor are proxies of type 3). The rest covariates age, weight index, and performance rating can be classified as type 1).

\textcolor{black}{Breaking the impasse of dependent right censoring in survival analysis, we introduce a novel proximal survival analysis framework with the help of these proxies, without stringent assumptions, like ``conditional independent censoring'', which rarely hold true in messy observational data. Instead of clinging to unattainable perfect measurement, proximal survival analysis embraces reality. We acknowledge the inherent difficulty of capturing all common factors and turn proxy variables, readily available in most datasets, into powerful allies. Our contribution stands as the first in survival analysis to leverage such an ambitious approach. While borrowing inspiration from ``proximal causal inference,'' past applications have been limited to discrete-time longitudinal studies. proximal survival analysis tackles a far more complex challenge, allowing continuous time for event occurrence, censoring, and time-varying covariates. This generalization presented monumental hurdles, both theoretical and computational, making our achievement a pivotal leap forward.} The proximal survival analysis' identification leverages three types of proxies to build an event-inducing bridge process and a censoring-inducing bridge process, which are solutions to some integral equations. These processes further yield a proximal event-inducing identification formula and a proximal censoring-inducing identification formula, respectively. These bridge processes together also provide a proximal doubly robust identification formula, which is shown to be doubly robust in the sense that it identifies the parameter of interest correctly provided that either the event-inducing bridge process or the censoring-inducing bridge process is correctly identified but not necessarily both. These three formulas suggest three adaptive estimators by estimating first the bridge processes and then the parameter of interest. The first-step estimators for bridge processes can be quite flexible and therefore we investigate generic assumptions under which the resulting estimators are consistent, asymptotically normal, and doubly robust. As an illustration, we consider a concrete setting where we only have baseline covariates but dependent censoring that cannot be explained away by these covariates. We build semiparametric models for two processes and construct estimators. We examine the finite-sample performance of these estimators via extensive Monte-Carlo simulations and apply them to the SEER-Medicare dataset. We note that our proximal survival analysis framework does not claim that one can handle unmeasured covariates for free, but rather offers an alternative method that practitioners can always consider if they have measured enough information though there are concerns that conditional independent censoring may not hold with the measured covariates exactly. \textcolor{black}{We deem that our proximal survival analysis framework offers a more plausible and potentially more preferable alternative for several reasons: firstly, it recognizes that perfectly measuring all common factors is often impossible. It allows us to proceed even with imperfect proxies, capturing the influence of unmeasured factors on both treatment and outcome. Unlike the strict non-informative censoring demanded by traditional methods, proximal survival analysis allows for some wiggle room. This makes it more applicable in messy observational data, where perfect control and measurement are elusive. Our framework utilizes the information contained in proxy variables, which we often have access to. This allows us to glean insights from data that might otherwise be discarded due to potential confounding.}

The remainder of the article is organized as follows. We give a review of the existing literature and methods for handling right censoring in Section \ref{sec:review}, where we also briefly review proximal causal inference. We give proximal assumptions and establish our proximal identification for survival analysis in the presence of dependent right censoring in Section \ref{sec:iden}, including three methods. We consider generic proximal estimation procedures and establish general asymptotic results in Section \ref{sec:est}. We consider plausible numerical setting under which our proximal assumption holds, propose concrete estimators, examine finite-sample performance of our estimators with simulations in Section \ref{sec:example}. We conduct a real data application using the publicly available SEER-Medicare dataset in Section \ref{sec:real}. We end the paper with a discussion in Section \ref{sec:dis}. 

\section{Review}\label{sec:review}

\subsection{A Review of Theory under Independent Censoring}
Suppose we observe a time to event variable subject to right censoring $\widetilde T = \min(T, C)$ and an event indicator $\Delta = \mathbbm{1}(T < C)$, where $T$ is time to event of interest and $C$ is right censoring time. The ``independent censoring'' assumption states that the event time and the censoring time are independent, that is,
\begin{equation}
    T \perp C.
\end{equation}
The Kaplan-Meier estimator, the Nelson-Aalen estimator, and the logrank test are typically used to address right censoring in this case.

\subsection{A Review of Theory under Conditional Independent Censoring}\label{sec:car}
However, when the ``independent censoring'' assumption fails, all aforementioned methods become invalid. Nonetheless, when possibly time-varying covariates information is given, practitioners may leverage it to handle dependent censoring. Conventionally this is accomplished by the ``conditional independent censoring'' assumption (or known as coarsened at random), which basically requires that the event time and the censoring time are independent within the covariate strata. Here we review identification under ``conditional independent censoring'' \citep{van2003unified, rotnitzky2005inverse, tsiatis2006semiparametric}. Let us denote possibly time-varying covariates as $L(t)$, where $\overline L(t) = \{L(s): 0 \leq s \leq t\}$. The observed data are $(\widetilde T, \Delta, \overline L(\widetilde T))$ and the full data are $(T, C, \overline L(T))$. The censoring time hazard is defined as
\begin{equation}
    \lambda_C^F\{t|T, \overline L(T)\} = \lim_{h \to 0} \frac{1}{h}\P\{C < t + h|C \geq t, T, \overline L(T)\}.
\end{equation}
The censoring mechanism is said to be conditionally independent if
\begin{equation}\label{eq:car}
    \lambda_C^F\{t|T, \overline L(T)\} = \lambda_C\{t|\overline L(t)\} := \lim_{h \to 0} \frac{1}{h}\P\{C < t + h|C \geq t, T \geq t, \overline L(t)\},
\end{equation}
where $:=$ means ``is defined as.'' That is, the instantaneous risk of censoring at time $t$ for individuals still at risk at time $t$ given their event time and all covariates is the same as the instantaneous risk of censoring at time $t$ for individuals still at risk at time $t$ given their history of covariates up to time $t$. This is to say that the instantaneous risk of censoring at time $t$ only depends on the history of covariates up to time $t$ and not on the future beyond $t$, which is quite a natural assumption. The proportional hazards model \citep{cox1972regression, cox1975partial} and the additive hazards model \citep{aalen1980model} are also well-studied models commonly used under ``conditional independent censoring''. \textcolor{black}{The conditional independent censoring condition \eqref{eq:car} can be equivalently understood as the ``coarsening at random'' assumption from the coarsened data (or missing data) literature \citep{heitjan1991ignorability, tsiatis2006semiparametric}, which is known to place no restrictions on the observed data distribution. This property justifies that the following identification is nonparametric.}

Suppose one is interested in identifying a finite-dimensional parameter $\theta$, which is a locally unique solution for some full data estimating equation as
\begin{equation}\label{eq:fullee}
    \E[D\{T, \overline L(T); \theta_0\}] = 0,
\end{equation}
at the truth $\theta_0$, where $D\{T, \overline L(T); \theta\}$ is the estimating function, and
\begin{equation}
    \inf_{\|\theta - \theta_0\|_1 > \eps}\left|\E[D\{T, \overline L(T); \theta\}]\right| > 0,
\end{equation}
for every $\eps > 0$, where $\|\theta - \theta_0\|_1 > \eps$ represents parameter subset outside of an $\eps$ neighborhood of $\theta_0$. For instance, $D\{T, \overline X(T); \theta\} = \mathbbm{1}(T > t) - \theta$ identifies the marginal survival probability. $D\{T, \overline X(T); \theta\} = \min(T, \tau) - \theta$ for some $\tau > 0$ identifies the restricted mean survival time. Under \eqref{eq:car} and additionally with positivity,
\begin{equation}\label{eq:pos}
    \P\{T \leq C|T, \bar L(T)\} > 0,
\end{equation}
almost surely, one can identify $\theta$ via the so-called inverse probability censoring weighting (IPCW) identification based on the observed data by solving 
\begin{equation}
    \E\left(\frac{\Delta D\{\widetilde T, \overline L(\widetilde T); \theta\}}{\Prodi_0^{\widetilde T} \left[1 - \lambda_C\left\{t|\overline L(t)\right\}dt\right]}\right) = 0,
\end{equation}
where $\prodi$ is the product integral \citep{gill1990survey, andersen2012statistical}. The IPCW identification can be augmented into the augmented inverse probability censoring weighting (AIPCW) as 
\begin{equation}\label{eq:aipcw}
    \E\left(\frac{\Delta D\{\widetilde T, \overline L(\widetilde T); \theta\}}{\Prodi_0^{\widetilde T} \left[1 - \lambda_C\left\{t|\overline L(t)\right\}dt\right]} + \int_0^{\widetilde T} \frac{\E\left[D\{T, \overline L(T); \theta\}|T \geq t, \overline L(t)\right\}}{\Prodi_0^t \left[1 - \lambda_C\left\{u|\overline L(u)\right\}du\right]}dM_C\{t, \overline L(t)\}\right) = 0,
\end{equation}
where
\begin{equation}
    dM_C(t) = \mathbbm{1}(t \leq \widetilde T < t + dt, \Delta = 0) - \mathbbm{1}(\widetilde T \geq t)\lambda_C\left\{t|\overline L(t)\right\}dt.
\end{equation}

\begin{rem}\label{rem:td1}
Define $T_D$ as the minimum time that $D\{T, \overline X(T); \theta\}$ is observed in full data, $\widetilde T_D = \min(T_D, C)$, and $\Delta_D = \mathbbm{1}(D \text{ is observed})$. For instance, when $D(T, \widetilde L(T); \theta) = \mathbbm{1}(T > t) - \theta$, $T_D = \min(T, t)$. Note that $T_D \leq T$ in the full data. Both $\widetilde T_D$, and $\Delta_D$ are observed. The additional notation $T_D$, $\widetilde T_D$, and $\Delta_D$ was employed \citep{van2003unified, rotnitzky2005inverse}. The resulting IPCW identification shall be a more efficient usage of data compared to \citet{tsiatis2006semiparametric} as
\begin{equation}
    \E\left(\frac{\Delta_D D\{\widetilde T_D, \overline L(\widetilde T_D); \theta\}}{\Prodi_0^{\widetilde T_D} \left[1 - \lambda_C\left\{t|\overline L(t)\right\}dt\right]}\right) = 0.
\end{equation}
The AIPCW becomes
\begin{equation}
    \E\left(\frac{\Delta_D D\{\widetilde T_D, \overline L(\widetilde T_D); \theta\}}{\Prodi_0^{\widetilde T_D} \left[1 - \lambda_C\left\{t|\overline L(t)\right\}dt\right]} + \int_0^{\widetilde T_D} \frac{\E\left[D\{T, \overline L(T); \theta\}|T \geq t, \overline L(t)\right]}{\Prodi_0^t \left[1 - \lambda_C\left\{u|\overline L(u)\right\}du\right]}dM_C\{t, \overline L(t)\}\right) = 0.
\end{equation}
When $T_D = T$, results in \citep{van2003unified, rotnitzky2005inverse} coincide \citet{tsiatis2006semiparametric}. Our results below share a similar property as here. That is, our results below remain to be correct when replacing $\widetilde T$ by $\widetilde T_D$ and $\Delta$ by $\Delta_D$. For ease of notation and better readability, we work with $\widetilde T$ and $\Delta$.
\end{rem}

\subsection{A Review when Conditional Independent Censoring Fails}\label{sec:carfails}
As discussed in the introduction, conditional independent censoring can be subject to questioning in practice. It is especially of concern under competing risks data if treating other failure types as conditional independent right censoring because individuals who experience competing events may have different prognoses for the primary event, which can hardly ever be captured by measured covariates. To handle this, alternative quantities such as subdistribution hazard \citep{beyersmann2008time, austin2020review}
\begin{equation}
    \lim_{h \to 0}\frac{1}{h}\P[t \leq \widetilde T < t + h, \Delta = 1|\{\widetilde T \geq t\} \cup \{\widetilde T \leq t, \Delta = 0\}, \overline L(t \wedge \widetilde T)],
\end{equation}
or cause-specific hazard \citep{kalbfleisch2011statistical},
\begin{equation}
    \lim_{h \to 0}\frac{1}{h}\P\{t \leq \widetilde T < t + h, \Delta = 1|\widetilde T \geq t, \overline L(t)\},
\end{equation}
are typically considered. However, both quantities involve the censoring variable (or in this case, competing events), which leaves the conventional quantities of interest in survival analysis like the survival function, hazard function, and restricted mean survival time \citep{royston2013restricted} unidentified. Quantities that are only defined by full data like in the form of \eqref{eq:fullee} could fail to be nonparametrically identified.

\subsection{\textcolor{black}{Proximal Causal Inference}}\label{sec:pci}
Here we give a brief review of the ``proximal causal inference'' identification framework for our readers who are unfamiliar with. The review is mostly borrowed from \citet{miao2018identifying, tchetgen2020introduction, cui2020semiparametric}. We define some temporary notation that is only used within this section. The results are stated with some minor adjustments. To better assist our readers to understand our framework, we will connect our assumptions and theorems to the ones listed below in parentheses.

Suppose we aim to estimate the effect of a binary treatment $A$ on an outcome $Y$ subject to potential unmeasured confounding even with observed covariates $L$. We let $U$ denote the unmeasured confounders, $Y(a)$, $a = 0, 1$ denote the potential outcome that would be observed if the treatment were set to $a$. For easier exposition, suppose we are interested in $\theta = \E[Y (1)]$, the average outcome if all patients were set to receive the treatment. We assume that the following ``consistency'' assumption holds: $Y = Y(A)$ almost surely. Suppose that one has partitioned the observed covariates $L$ into variables $(X,Z,W)$, such that $Z$ includes treatment-inducing confounding proxies, and $W$ includes outcome-inducing confounding proxies, which satisfy the ``proximal independence'' assumption (Assumption \ref{assump:latentcar}):
\begin{equation}
    (W, Y (a)) \perp A|U,X, Z,
\end{equation}
\begin{equation}
    (W, Y (a)) \perp Z|U,X ~\text{for}~ a=0,1,
\end{equation}
the positivity assumption (Assumption \ref{assump:proximalpos}): $0<\Pr(A=a|U,X)<1$ almost surely, $a=0,1$, and the completeness assumption (Assumption \ref{assump:eventuntestcomplete}): for any square-integrable function $g$ and for any $a,x$, $\E[g(U)|Z, A=a, X=x]=0$ almost surely if and only if $g(U)=0$ almost surely. Furthermore, suppose that there exists an outcome confounding bridge function $h(w, a, x)$ that solves the following integral equation (Assumption \ref{assump:eventbridge}):
\begin{equation}\label{eq:condexpec}
    \E[Y - h(W, A, X)|Z,A,X] = 0,
\end{equation}
almost surely, this implies
\begin{equation}
    \E[Y - h(W, A, X)|U,A,X] = 0,
\end{equation}
almost surely, and hence 
\begin{equation}
    \E[Y(1)] = E\{h(W, 1, X)\}.
\end{equation}
This is known as the ``proximal g-formula'' for identification (Theorem \ref{thm:eventiden}). Another identification is called proximal inverse probability weighting (Theorem \ref{thm:censoringiden}). We impose another completeness assumption (Assumption \ref{assump:censoruntestcomplete}): For any square-integrable function $g$ and for any $a,x$, $\E[g(U)|W, A=a, X=x]=0$ almost surely if and only if $g(U)=0$ almost surely. Then suppose that there exists a treatment confounding bridge function $q(z,a,x)$ that solves the integral equation (Assumption \ref{assump:censoringbridge}):
\begin{align}\label{eq:condexpect2}
\E[q(Z,a,X)|W,A=a, X] = \frac{1}{f(A=a|W, X)},
\end{align}
almost surely, this implies 
\begin{equation}
    \E[q(Z,a,X)|U,A=a, X] = \frac{1}{f(A=a|U,X)},
\end{equation}
almost surely. Furthermore, one also has that
\begin{equation}
    \E\{Y(1)\} = \E\{I(A=1)Yq(Z,1,X)\} ,
\end{equation}
Finally, there is a doubly robust identification (Theorem \ref{thm:driden}):
\begin{align}
\Xi(h, q; \theta) = I(A=1)q(Z,A,X) [Y- h(W,A,X)] + h(W,1,X).
\end{align}
which admits the double robustness property: $\E\{Y(1)\} = \E[\Xi(h_{*}, q_{*};\theta)]$ provided that either $h_{*}$ is a solution to equation \eqref{eq:condexpec}, or $q_{*}$ is a solution to equation \eqref{eq:condexpect2}, but both do not necessarily hold. 

\section{Proximal Identification}\label{sec:iden}
Suppose we observe $\{\widetilde T = \min(T, C), \Delta = \mathbbm{1}(T \leq C), \overline L(\widetilde T)\}$, where $T$ and $C$ are the underlying true event time and censoring time. We consider the scenario under which \eqref{eq:car} fails to hold with the measured covariate $L(t)$, owing to the presence of some unmeasured covariates $U(t)$. \textcolor{black}{Both $L(t)$ and $U(t)$ are allowed to be defined until $\infty$, denoting end of the study. A positive constant can be used to replace $\infty$ in practice. We use $\infty$ to fully generalize \citep{rotnitzky2005inverse, tsiatis2006semiparametric}.} In the following, we propose to replace \eqref{eq:car} with an assumption that an analyst has correctly identified the observed covariates $L(t)$ into three types of proxies: \textcolor{black}{type 1) proxies $X(t)$, type 2) proxies $Z(t)$, and type 3) proxies $W(t)$}, as mentioned in the introduction, which formally satisfy:
\begin{assump}[Proximal Independence]\label{assump:latentcar}
\begin{equation}
    \overline Z(t) \perp \overline W(t)~|~C \geq t, T \geq t, \overline U(t), \overline X(t),
\end{equation}
and
\begin{equation}
    \lambda_C^F\{t|T, \overline U(T), \overline X(T), \overline W(T), \overline Z(T)\} = \lim_{h \to 0} \frac{1}{h}\P\{C < t + h|C \geq t, T \geq t, \overline U(t), \overline X(t), \overline Z(t)\},
\end{equation}
where $\lambda_C^F(t|\cdot)$ is defined as
\begin{equation}
    \lambda_C^F(t|\cdot):= \lim_{h \to 0} \frac{1}{h}\P(C < t + h|C \geq t, \cdot).
\end{equation}
\end{assump}
A few comments are in order. Firstly, this assumption formally encodes that \eqref{eq:car} holds by adjusting for $X(t), U(t)$, that is,
\begin{equation}
    \lim_{h \to 0} \frac{1}{h}\P\{C < t + h|C \geq t, T, \overline U(T), \overline X(T)\} = \lim_{h \to 0} \frac{1}{h}\P\{C < t + h|C \geq t, T \geq t, \overline U(t), \overline X(t)\}.
\end{equation}
This latent ``conditional independent censoring'' assumption is plausible provided that $U(t)$ are sufficiently enriched to include all underlying associations between $T$ and $C$ not captured by $X(t)$. As $U(t)$ is not required to be observed, this assumption will generally hold. Secondly, $\{T, \overline W(T)\}$ and $\{\mathbbm{1}(t \leq C < t + dt), \overline Z(t)\}$ are allowed to be dependent but this dependency totally explained by $\overline U(t), \overline X(t)$. As defined, $\overline Z(t)$ are only associated with $T$ through $\overline U(t), \overline X(t)$, so are $\overline W(t)$ and $C$. To ensure that there are enough data for identification, we also assume that
\begin{assump}[Latent Positivity]\label{assump:proximalpos}
The conditional probability of not being censored on $T, \overline U(T), \overline X(T)$ is positive, that is,
\begin{equation}\label{eq:proximalpos}
    \P\{\Delta = 1|T, \overline U(T), \overline X(T)\} > 0,
\end{equation}
almost surely.
\end{assump}
\textcolor{black}{This assumption is a latent version of \eqref{eq:pos}, which hinges on the unobserved process $\bar U(T)$. It reflects a belief that for any possible unknown factors $\bar U(T)$, possibly prognosis, the chance of experiencing events instead of being censored is positive. Note that by Assumption \ref{assump:latentcar}, }
\begin{equation}
    \P\{\Delta = 1|T, \overline U(T), \overline X(T)\} = \Prodi_0^T[1 - \lambda_C^F\{t|\overline U(t), \overline X(t)\}dt].
\end{equation}

As we mentioned in Section \ref{sec:car}, scientific interest typically focuses on a finite-dimensional parameter $\theta$ as some functions of the full data distribution, which we assume can be identified by a full data estimating equation of the same dimension
\begin{equation}\label{eq:fulldataX}
    D\{T, \overline X(T); \theta\}.
\end{equation}
where $D$ is a known function. That is,
\begin{equation}
    \E[D\{T, \overline X(T); \theta_0\}] = 0,
\end{equation}
at the truth $\theta_0$ and
\begin{equation}
    \inf_{\|\theta - \theta_0\|_1 > \eps}\left|\E[D\{T, \overline X(T); \theta\}]\right| > 0,
\end{equation}
for every $\eps > 0$, where $\|\theta - \theta_0\|_1 > \eps$ represents parameter subset outside of an $\eps$ neighborhood of $\theta_0$. However, in this time, we no longer base the identification framework on \eqref{eq:car}. 

\subsection{Identification via Event-inducing bridge process}
Adopting the common counting process notation \citep{andersen2012statistical}, we define $dN_T(t) = \mathbbm{1}(t \leq \widetilde T < t + dt, \Delta = 1) = \mathbbm{1}(t \leq T < t + dt, T < C)$. For any function $f$ over time $t$ we define
\begin{equation}
    V_0^\infty(f) = \sup_{\cP}\sum_{m = 0}^M|f(t_{m + 1}) - f(t_m)|,
\end{equation}
where the supremum is over $\cP =\left\{P=\{t_{0},\cdots, t_{M}\}\mid P{\text{ is a  partition of }}[0,\infty)\right\}$, the set of all possible partitions of the given interval, . We say a function $f(t)$ over time $t$ of bounded variation if $V_0^\infty(f) < \infty$. We call a stochastic process $\cO(t)$ of pathwise bounded variation if its paths are of bounded variation almost surely. We impose the following assumption on the observed data to guarantee the existence of an event-inducing bridge process (defined in the assumption), which proves central to our identification framework.
\begin{assump}[Existence of an Event-inducing Bridge Process]\label{assump:eventbridge}
There exists an {\it event-inducing bridge process} $h_0\{t, \overline w(t), \overline x(t); D, \theta\}$ of pathwise bounded variation satisfying
\begin{equation}\label{eq:eventconfbridgeiden}
    \E[dH_0(t; D, \theta) - H_0(t; D, \theta)dN_T(t)|\widetilde T \geq t, \overline Z(t), \overline X(t)] = 0,
\end{equation}
where 
\begin{equation}
    H_0(t; D, \theta) = h_0\{t, \overline W(t), \overline X(t); D, \theta\},
\end{equation}
with an initial condition
\begin{equation}\label{eq:eventconfbridgeideninit}
    h_0\{\infty, t, \overline x(t); D, \theta\} = D\{t, \overline x(t); \theta\}.
\end{equation}
\end{assump}
We take the existence of an event-inducing bridge process as a primitive assumption. In the case of finite-dimensional random variables \citep{carrasco2007linear, darolles2011nonparametric, miao2018identifying, ying2021proximal} instead of infinite-dimensional random processes, \eqref{eq:eventconfbridgeiden} is a Fredholm integral equation. By Picard's theorem \citep{kress1989linear}, the existence of such an integral equation is equivalent to some properties of the kernel in the integral equation. A study of the existence to such a process is of probabilistic interest but beyond statistical interest in this paper. It is well-known such an integral equation is ill-posed and thus might have multiple solutions. Note that uniqueness is not essential since any solution can identify $\theta$ later. \textcolor{black}{To demystify it for our draft, we provide a parametric example in Section \ref{sec:compatible} where Assumption \ref{assump:eventbridge} can be shown to hold. }

The Equation \eqref{eq:eventconfbridgeiden} with an initial condition \eqref{eq:eventconfbridgeideninit} can equivalently be understood as requiring
\begin{equation}\label{eq:eventconfbridgeidenmargin}
    \E\left[\int_0^{\widetilde T}q'\{t, \overline Z(t), \overline X(t)\}\{dH_0(t; D, \theta) - H_0(t; D, \theta)dN_T(t)\}\right] = 0,
\end{equation}
for any function $q'\{t, \overline z(t), \overline x(t)\}$, where the integration is in the Riemann-Stietjes sense. We further require the following assumption on the full data for valid survival interpretation of \eqref{eq:eventbridgeiden} later.
\begin{assump}[Proxy Relevance for Event-inducing Bridge Process]\label{assump:eventuntestcomplete}
For any $t, \overline X(t)$, and any integrable function $\nu(\cdot)$,
\begin{equation}\label{eq:eventcompleteuntest}
    \E[\nu\{\overline U(t)\}|\widetilde T \geq t, \overline Z(t), \overline X(t)] = 0~\text{if and only if}~\nu\{\overline U(t)\} = 0~\text{almost surely},
\end{equation}
\end{assump}
This condition is formally known as the completeness condition which can accommodate categorical, discrete, and continuous variables. Completeness is essential to ensure the identification of $\theta$. Here one may interpret the completeness condition \eqref{eq:eventcompleteuntest} as a requirement relating the range of $\overline U(t)$ to that of $\overline Z(t)$ which essentially states that the set of proxies must have sufficient variability relative to the variability of $\overline U(t)$ within the risk set.
Completeness is a familiar technical condition central to the study of sufficiency in the foundational theory of statistical inference. Many commonly-used parametric and semiparametric models such as the semiparametric exponential family \citep{newey2003instrumental} and semiparametric location-scale family \citep{hu2018nonparametric} satisfy the completeness condition. For nonparametric regression models, results of \cite{d2011completeness} and \cite{darolles2011nonparametric} can be used to justify the completeness condition, although their primary focus is on a nonparametric instrumental variable model, where completeness plays a central role. See \cite{chen2014local}, \cite{andrews2017examples}, \citet{ying2021proximal}, and references therein for an excellent overview of the role of completeness. With assumptions above prepared, we have:
\begin{thm}[Proximal Event-inducing Identification]\label{thm:eventiden}
Under Assumptions \ref{assump:latentcar}, \ref{assump:proximalpos}, \ref{assump:eventbridge}, and \ref{assump:eventuntestcomplete}, for any event-inducing bridge process $H_0(t; D, \theta) = h_0\{t, \overline W(t), \overline X(t); D, \theta\}$ satisfying \eqref{eq:eventconfbridgeiden} and \eqref{eq:eventconfbridgeideninit}, it also satisfies
\begin{equation}\label{eq:eventbridgeU}
    \E\{dH_0(t; D, \theta) - H_0(t; D, \theta)dN_T(t)|\widetilde T \geq t, \overline U(t), \overline X(t)\} = 0,
\end{equation}
with an initial condition
\begin{equation}\label{eq:eventbridgeUinit}
    h_0\{\infty, t, \overline x(t); D, \theta\} = D\{t, \overline x(t); \theta\}.
\end{equation}
Furthermore, we have
\begin{equation}\label{eq:eventbridgeiden}
    \E[D\{T, \overline X(T); \theta\}] = \E\{H_0(0; D, \theta)\},
\end{equation}
for any $\theta$.
\end{thm}

When ``conditional independent censoring'' \eqref{eq:car} actually holds as in Section \ref{sec:car} given $L(t) = (X(t), W(t))$, such that $Z(t) = W(t)$, we show that
\begin{equation}\label{eq:carh}
    H_0(t; D, \theta) = \E[D\{T, \overline L(T); \theta\}|T \geq t, \overline L(t)],
\end{equation}
satisfies \eqref{eq:eventbridgeU} and \eqref{eq:eventbridgeUinit}. This can be shown by incorporating martingale theory and Doob's Theorem, which we omit here. 
Therefore \eqref{eq:eventbridgeiden} becomes
\begin{equation}
    \E[D\{T, \overline X(T); \theta\}] = \E(\E[D\{T, \overline L(T); \theta\}|T \geq 0, \overline L(0)]),
\end{equation}
when conditional independent censoring happens to hold.

\subsection{Identification via Censoring-inducing Bridge Process}
Define $dN_C(t) = \mathbbm{1}(t \leq \widetilde T < t + dt, \Delta = 0) = \mathbbm{1}(t \leq C < t + dt, C < T)$.
\begin{assump}[Existence of a Censoring-inducing Bridge Process]\label{assump:censoringbridge}
There exists a {\it censoring-inducing bridge process} $q_0\{t, \overline z(t), \overline x(t)\}$ of pathwise bounded variation satisfying
\begin{equation}\label{eq:censorconfbridgeiden}
    \E\{dQ_0(t) - Q_0(t)dN_C(t)|\widetilde T \geq t, \overline W(t), \overline X(t)\} = 0,
\end{equation}
where 
\begin{equation}
    Q_0(t) = q_0\{t, \overline Z(t), \overline X(t)\},
\end{equation}
with an initial condition
\begin{equation}\label{eq:censorconfbridgeideninit}
    q_0\{0, \overline z(0), \overline x(0)\} = 1.
\end{equation}
\end{assump}
The differential equation \eqref{eq:censorconfbridgeiden} with an initial condition \eqref{eq:censorconfbridgeideninit} can be equivalently understood as claiming 
\begin{equation}\label{eq:censorconfbridgeidenmargin}
    \E\left[\int_0^{\widetilde T}h'\{t, \overline w(t), \overline x(t)\}\{dQ_0(t) - Q_0(t)dN_C(t)\}\right] = 0,
\end{equation}
for any function $h'\{t, \widetilde w(t), \widetilde x(t)\}$, where the integral is pathwise Riemann-Stietjes integral and
\begin{equation}
    Q_0(0) = 1,
\end{equation}
for any measurable function $h'\{t, \overline W(t), \overline X(t)\}$. Interestingly, by Assumption \ref{assump:eventbridge} one learns $H_t$ from the farthest away at $\widetilde T$ and back to time zero, whilst Assumption \ref{assump:censoringbridge} here implies one learns $Q_t$ from time zero to the farthest away.

\begin{assump}[Sequential Proxy Relevance for Censoring-inducing Bridge Process]\label{assump:censoruntestcomplete}
For any $t, \overline W(t), \overline X(t)$, and any integrable function $\nu(\cdot)$,
\begin{equation}\label{eq:censorcompleteuntest}
    \E[\nu\{\overline U(t)\}|\widetilde T \geq t, \overline W(t), \overline X(t)] = 0~\text{if and only if}~\nu\{\overline U(t)\} = 0~\text{almost surely},
\end{equation}
\end{assump}

\begin{thm}[Proximal Censoring-inducing Identification]\label{thm:censoringiden}
Under Assumptions \ref{assump:latentcar}, \ref{assump:proximalpos}, \ref{assump:censoringbridge} and \ref{assump:censoruntestcomplete}, for any censoring-inducing bridge process $Q_0(t) = q_0\{t, \overline Z(t), \overline X(t)\}$ satisfying \eqref{eq:censorconfbridgeiden} and \eqref{eq:censorconfbridgeideninit}, it also satisfies
\begin{equation}\label{eq:censorbridgeU}
    \E\{dQ_0(t) - Q_0(t)dN_C(t)|\widetilde T \geq t, \overline U(t), \overline X(t)\} = 0,
\end{equation}
with an initial condition
\begin{equation}\label{eq:censorbridgeUinit}
    q_0\{0, \overline z(0), \overline x(0)\} = 1.
\end{equation}
Furthermore, we have
\begin{equation}\label{eq:censorbridgeiden}
    \E[D\{T, \overline X(T); \theta\}] = \E[\Delta Q_0(\widetilde T) D\{\widetilde T, \overline X(\widetilde T); \theta\}],
\end{equation}
for any $\theta$.
\end{thm}

When conditional independent censoring actually holds as in Section \ref{sec:car} given $L(t) = (X(t), Z(t))$, such that $W(t) = Z(t)$, we can show that
\begin{equation}
    Q_0(t) = \frac{1}{\prodi_0^t[1 - \mathbbm{1}(\widetilde T \geq t)\lambda_C\{s|\overline L(s)\}ds]},
\end{equation}
satisfies \eqref{eq:censorbridgeU} and \eqref{eq:censorbridgeUinit}. This is because \eqref{eq:censorbridgeU} becomes
\begin{align}
    &\E\left\{dQ_0(t) - Q_0(t)dN_C(t)|\widetilde T \geq t, \overline L(t)\right\} \\
    &= \E\left(\frac{\lambda_C(t|\overline L)dt}{\prodi_0^t[1 - \lambda_C\{s|\overline L(s)\}ds]} - \frac{dN_C(t)}{\prodi_0^t[1 - \lambda_C\{s|\overline L(s)\}ds]}\Bigg|\widetilde T \geq t, \overline L(t)\right)\\
    &= \frac{1}{\prodi_0^t[1 - \lambda_C\{s|\overline L(s)\}ds]}\E\left[\lambda_C\{t|\overline L(t)\}dt - dN_C(t)|\widetilde T \geq t, \overline L(t)\right]\\
    &= 0,
\end{align}
by the definition of $\lambda_C(t|\overline L)$. The Equation \eqref{eq:censorbridgeUinit} is trivially satisfied. Therefore \eqref{eq:censorbridgeiden} becomes
\begin{equation}
    \E[D\{T, \overline X(T); \theta\}] = \E\left(\frac{\Delta D\{\widetilde T, \overline X(\widetilde T); \theta\}}{\prodi_0^{\widetilde T}[1 - \lambda_C\{s|\overline L(s)\}ds]}\right),
\end{equation}
when conditional independent censoring happens to hold.

Other than \eqref{eq:eventbridgeiden} and \eqref{eq:censorbridgeiden}, there is a third identifying method leveraging both $H_0$ and $Q_0$ that proves to provide extra protection against process mis-specification, stated as follows.
\begin{thm}[Proximal Doubly Robust Identification]\label{thm:driden}
Under Assumptions \ref{assump:latentcar} - \ref{assump:censoringbridge}, if $H_0(t; D, \theta) = h_0\{t, \overline W(t), \overline X(t); D, \theta\}$ satisfies \eqref{eq:eventconfbridgeiden} and \eqref{eq:eventconfbridgeideninit} and $Q_0(t) = q_0\{t, \overline Z(t), \overline X(t)\}$ satisfies \eqref{eq:censorconfbridgeiden} and \eqref{eq:censorconfbridgeideninit}, we have
\begin{equation}
    \E[D\{T, \overline X(T); \theta\}] = \E\{\Xi(H_0, Q; D, \theta)\} = \E\{\Xi(H, Q_0; D, \theta)\},
\end{equation}
where 
\begin{equation}\label{eq:xi}
    \Xi(H, Q; D, \theta) 
    := \Delta Q(\widetilde T) D\{\widetilde T, \overline X(\widetilde T); \theta\} - \int_0^{\widetilde T}H(t; D, \theta)\{dQ(t) - dN_C(t)Q(t)\},
\end{equation}
for any $H(t; D, \theta)$ and $Q(t)$ whenever the Riemann-Stietjes integral involved in \eqref{eq:xi} is well-defined.
\end{thm}
The theorem states that the observed data estimating equation $\E\{\Xi(H_0, Q; D, \theta)\}$ equals $\E[D\{T, \overline X(T); \theta\}]$ provided that either $H(t; D, \theta)$ or $Q(t)$ is correctly specified but not necessarily both. Note that with integration by parts,
\begin{align}
    \Xi(H, Q; D, \theta) = \int_0^{\widetilde T}Q(t)\{dH(t; D, \theta) - H(t; D, \theta)dN_T(t)\} + H(0; D, \theta).
\end{align}
An important observation is that 
\begin{equation}
    \Xi(H, 0; D, \theta) = H(0; D, \theta),
\end{equation}
in \eqref{eq:eventbridgeiden} and
\begin{equation}
    \Xi(0, Q; D, \theta) = \Delta Q(\widetilde T) D\{\widetilde T, \overline X(\widetilde T); \theta\},
\end{equation}
in \eqref{eq:censorbridgeiden}.
\textcolor{black}{In Section \ref{sec:compatibility}, we prove that Assumptions \ref{assump:latentcar}-\ref{assump:censoruntestcomplete} all hold under the simulation setting in Section \ref{sec:simu}. }
\begin{rem}\label{rem:td2}
As in Remark \ref{rem:td1}, our results remain correct if replacing $(\widetilde T, \Delta)$ by $(\widetilde T_D, \Delta_D)$ throughout this section, where $T_D$ is the minimum time that $D\{T, \overline X(T); \theta\}$ is observed in full data, $\widetilde T_D = \min(T_D, C)$, and $\Delta_D = \mathbbm{1}(D \text{ is observed})$. For example, \eqref{eq:xi} becomes
\begin{equation}\label{eq:xi2}
    \Xi(H, Q; D, \theta) 
    := \Delta_D Q(\widetilde T_D) D\{\widetilde T, \overline X(\widetilde T); \theta\} - \int_0^{\widetilde T_D}H(t; D, \theta)\{dQ(t) - dN_C(t)Q(t)\}.
\end{equation}
This notation though complicates the notation but proves useful in the simulation later in Section \ref{sec:example}.
\end{rem}

When conditional independent censoring actually holds as in Section \ref{sec:car} given $L(t) = (X(t), Z(t))$, such that $W(t) = Z(t)$, \eqref{eq:xi} becomes \eqref{eq:aipcw}.

\section{Proximal Estimation and Inference}\label{sec:est}
Suppose we observe $n$ i.i.d. samples $\{\widetilde T_i, \Delta_i, \overline X_i(\widetilde T_i), \overline W_i(\widetilde T_i), \overline Z_i(\widetilde T_i)\}$ and for convenience we write
\begin{equation}
    H_i(t; D, \theta) = h\{t, \overline W_i(t), \overline X_i(t); D, \theta\},
\end{equation}
\begin{equation}
    Q_i(t) = q\{t, \overline Z_i(t), \overline X_i(t)\},
\end{equation}
and
\begin{equation}
    \Xi_i(H, Q; D, \theta) = \Delta_i Q_i(T_i)D\{T_i, \overline X_i(T_i); \theta\} - \int_0^{\widetilde T_i}H_i(t; D, \theta)\{dQ_i(t) - Q_i(t)dN_{C, i}(t)\}.
\end{equation}
It is straightforward to see that $\widehat{\theta}$ can be obtained by solving $\frac{1}{n}\sum_{i = 1}^n\Xi_i(H_0, Q_0; D, \theta) = o_p(n^{-\frac{1}{2}})$ (or $\frac{1}{n}\sum_{i = 1}^n\Xi_i(H_0, 0; D, \theta) = o_p(n^{-\frac{1}{2}})$, $\frac{1}{n}\sum_{i = 1}^n\Xi_i(0, Q_0; D, \theta) = o_p(n^{-\frac{1}{2}})$); however, clearly such an estimator is technically not feasible as it depends crucially on complicated unknown processes $h_0$ and $q_0$ that satisfy Assumptions \ref{assump:eventbridge} and \ref{assump:censoringbridge}. Therefore, for practical purposes, we consider adaptive estimators that estimate $h_0\{t, w(t), x(t); D, \theta\}$, $q_0\{t, z(t), x(t)\}$  in the first step and then solve $\frac{1}{n}\sum_{i = 1}^n\Xi_i(\hat H, \hat Q; D, \theta) = o_p(n^{-\frac{1}{2}})$ in the second step. 

There are many possible approaches. For instance, one may consider first imposing compatible semiparametric working models \citep{tchetgen2020introduction, cui2020semiparametric, ying2021proximal, ying2022proximal} on $h$ and $q$. This suggests parameterizing $h$ and $q$ with possibly time-varying parameters $B(t)$ and $A(t)$, possibly as
\begin{equation}
    H\{t; D, \theta, B(t)\} = h\{t, \overline W(t), \overline X(t); D, \theta, B(t)\},
\end{equation}
\begin{equation}
    Q\{t; A(t)\} = q\{t, \overline Z(t), \overline X(t); A(t)\}.
\end{equation}
Then one may choose appropriate $q'$ and $h'$ in the following the estimating equations
\begin{equation}\label{eq:hmarginsample}
    \frac{1}{n}\sum_{n = 1}^n\left(\int_0^{\widetilde T_i}q'\{t, \overline Z_i(t), \overline X_i(t)\}[dH_i\{t; D, \theta, B(t)\} - H_i\{t; D, \theta, B(t)\}dN_{T, i}(t)]\right) = o_p(n^{-\frac{1}{2}}),
\end{equation}
\begin{equation}\label{eq:hmargininitsample}
    \frac{1}{n}\sum_{n = 1}^n\left(\Delta_i q'\{\widetilde T_i, \overline Z_i(\widetilde T_i), \overline X_i(\widetilde T_i)\}[D\{\widetilde T_i, \overline X_i(\widetilde T_i); \theta\} - H_i\{\widetilde T_i; D, \theta, B(\widetilde T_i)\}]\right) = o_p(n^{-\frac{1}{2}}),
\end{equation}
\begin{equation}\label{eq:qmarginsample}
    \frac{1}{n}\sum_{n = 1}^n\left(\int_0^{\widetilde T_i}h'\{t, \overline W_i(t), \overline X_i(t)\}[dQ_i\{t; A(t)\} - Q_i\{t; A(t)\}dN_{C, i}(t)]\right) = o_p(n^{-\frac{1}{2}}),
\end{equation}
\begin{equation}\label{eq:qmargininitsample}
    Q_i\{0; A(0)\} = 0,
\end{equation}
suggested by \eqref{eq:eventconfbridgeidenmargin} and \eqref{eq:censorconfbridgeidenmargin}, to solve for $B(t)$ and $A(t)$. In this way, commonly root-$n$ consistent and asymptotically normal estimators can be obtained for $B(t)$ and $A(t)$ under some regularity conditions, which lead to root-$n$ consistent and asymptotically normal estimators for $\theta$ as well. 

We provide generic sufficient assumptions on first-step estimators upon which the resulting estimator $\hat \theta$ are consistent and asymptotically normal for $\theta_0$. Then we construct estimators satisfying these assumptions to illustrate our framework in a concrete example in Section \ref{sec:example}. We assume based on i.i.d. samples, one has constructed estimators for $h_0$ and $q_0$:
\begin{equation}
    \hat h\{t, \overline w(t), \overline x(t); D, \theta\},
\end{equation}
for each fixed $\theta$, and
\begin{equation}
    \hat q\{t, \overline z(t), \overline x(t)\}.
\end{equation}
Then we define the {\it proximal event-inducing estimator} $\hat \theta_{\text{PEE}}$ as the solution to 
\begin{equation}
    \frac{1}{n}\sum_{i = 1}^n \hat H_i(0; D, \theta) = o_p(n^{-\frac{1}{2}}),
\end{equation}
the {\it proximal censoring-inducing estimator} $\hat \theta_{\text{PCE}}$ as the solution to
\begin{equation}
    \frac{1}{n}\sum_{i = 1}^n \Delta_i\hat Q_i(T_i)D\{T_i, \overline X_i(T_i); \theta\} = o_p(n^{-\frac{1}{2}}),
\end{equation}
the {\it proximal doubly robust estimator} $\hat \theta_{\text{PDRE}}$ as the solution to 
\begin{align}
    \frac{1}{n}\sum_{i = 1}^n \Xi_i(\hat H, \hat Q; D, \theta) = o_p(n^{-\frac{1}{2}}).
\end{align}
Below we provide assumptions upon which $\hat \theta_{\text{PEE}}$, $\hat \theta_{\text{PCE}}$, and $\hat \theta_{\text{PDRE}}$ are consistent and asymptotically normal estimators for $\theta_0$.

\subsection{Asymptotic Inference}
For any stochastic process $\cO(t)$, define the $L_2$-norm
\begin{equation}
    \|\cO(t)\|_2 = \E\{\cO(t)^2\}^\frac{1}{2},
\end{equation}
the $L_2$ supremum norm
\begin{equation}
    \|\cO(t)\|_{\sup, 2} = \|\sup_t|\cO(t)|\|_2,
\end{equation}
the $L_2$ total variation norm
\begin{equation}
    \|\cO(t)\|_{\text{TV}, 2} = \|V_0^\infty(\cO)\|_2.
\end{equation}
We assume the build estimators for processes satisfy the following conditions:
\begin{assump}[Uniform Convergence]\label{assump:uniformcons1}
~~~

\begin{enumerate}
    \item We have
\begin{equation}
    \left\|\hat H(t; D, \theta)\right\|_{\text{TV}, 2} < \infty,
\end{equation}
\begin{equation}
    \left\|\hat H(t; D, \theta) - H_*(t; D, \theta)\right\|_{\sup, 2} = o(1),
\end{equation}
for some stochastic process $H_*(t; D, \theta) = h_*\{t, \overline W(t), \overline X(t); D, \theta\}$ with finite $L_2$ total variation norm, that is, $\left\|H_*(t; D, \theta)\right\|_{\text{TV}, 2} < \infty$.
    \item We have
\begin{equation}
    \left\|\hat Q(t)\right\|_{\text{TV}, 2} < \infty,
\end{equation}
\begin{equation}
    \left\|\hat Q(t) - Q_*(t)\right\|_{\sup, 2} = o(1),
\end{equation}
for some stochastic process $Q_*(t) = q_*\{t, \overline Z(t), \overline X(t)\}$ with finite $L_2$ total variation norm, that is, $\left\|Q_*(t)\right\|_{\text{TV}, 2} < \infty$.
\end{enumerate}
\end{assump}
The uniform convergence assumption requires that our estimators converge uniformly to a limit process in probability. It is usually satisfied with estimators based on Gilvenko-Cantelli class \citep{van1997weak}. The following asymptotic linearity assumption is needed to ensure asymptotic normality.
\begin{assump}[Asymptotic Linearity]\label{assump:if}
~~~

\begin{enumerate}
    \item We have
\begin{equation}
    \left\|\hat H(t; D, \theta) - H_*(t; D, \theta) - \frac{1}{n}\sum_{i = 1}^n \xi_{h, i}\{t, \overline W(t), \overline X(t)\}\right\|_{\sup, 2} = o(n^{-\frac{1}{2}}),
\end{equation}
for some stochastic process $H_*(t; D, \theta) = h_*\{t, \overline W(t), \overline X(t); D, \theta\}$ and some stochastic process $\xi_h\{t, \overline w(t), \overline x(t)\}$, where $\E[\xi_h\{t, \overline w(t), \overline x(t)\}] = 0$;
    \item We have
\begin{equation}
    \left\|\hat Q(t) - Q_*(t) - \frac{1}{n}\sum_{i = 1}^n \xi_{q, i}\{t, \overline Z(t), \overline X(t)\}\right\|_{\sup, 2} = o(n^{-\frac{1}{2}}),
\end{equation}
for some stochastic process $Q_*(t) = q_*\{t, \overline Z(t), \overline X(t)\}$ and some stochastic process $\xi_q\{t, \overline z(t), \overline x(t)\}$, where $\E[\xi_q\{t, \overline z(t), \overline x(t)\}] = 0$;
    \item 
    We have either
\begin{equation}
    \left\|\hat H(t; D, \theta) - H_*(t; D, \theta)\right\|_{\sup, 2} = O(n^{-\frac{1}{2}}),
\end{equation}
\begin{equation}
    \left\|\hat H(t; D, \theta) - H_*(t; D, \theta) - \frac{1}{n}\sum_{i = 1}^n \xi_{h, i}\{t, \overline W(t), \overline X(t)\}\right\|_{\text{TV}, 2} = o(n^{-\frac{1}{2}}),
\end{equation}
or
\begin{equation}
    \left\|\hat Q(t) - Q_*(t)\right\|_{\sup, 2} = O(n^{-\frac{1}{2}}),
\end{equation}
\begin{equation}
    \left\|\hat Q(t) - Q_*(t) - \frac{1}{n}\sum_{i = 1}^n \xi_{q, i}\{t, \overline Z(t), \overline X(t)\}\right\|_{\text{TV}, 2} = o(n^{-\frac{1}{2}}).
\end{equation}
\end{enumerate}

\end{assump}
The asymptotic linearity assumption requires that our estimators have asymptotically linear expansion in the $L_2$ supremum norm. Furthermore, one of the asymptotically linear expansions can be strengthened to $L_2$ total variation norm and the corresponding estimators converge in root-$n$ rate. It is usually satisfied with estimators based on Donsker class \citep{van1997weak}. This type of asymptotic linearity in the total variation sense was also considered in \citet{hou2021treatment, wang2022doubly}, and is intensively investigated in \citet{ying2023cautionary}. \textcolor{black}{In general, the rate conditions in terms of $\|\cdot\|_{\text{TV},2}$ does not imply those under $\|\cdot\|_{\sup,2}$, and vice versa. As an example, if $\hat F$ is the empirical cumulative distribution function and $F$ is the true cumulative distribution function, $\|\hat F - F\|_{\sup, 2}$ is usually of order $O(n^{-1/2})$, but $\|\hat F - F\|_{\text{TV},2}$ is $O(1)$. On the other hand, the rate conditions on the remainder terms in Assumption \ref{assump:if} are reasonable; for instance, \citet{wang2022doubly} have shown that estimated conditional cumulative hazard functions output from Cox regression and Aalen regression satisfy Assumptions \ref{assump:uniformcons1}, \ref{assump:if}. In Section \ref{sec:compatibility}, we show that our concrete estimators considered in Section \ref{sec:concrete} under the simulation setting in Section \ref{sec:simu} satisfy Assumptions \ref{assump:uniformcons1}, \ref{assump:if}.}

\begin{thm}[Consistency]\label{thm:cons}
~~~
\begin{enumerate}
    \item Under Assumptions \ref{assump:latentcar}, \ref{assump:proximalpos}, \ref{assump:eventbridge}, \ref{assump:eventuntestcomplete}, and \ref{assump:uniformcons1}(a) with $h_* = h_0$, $|\hat \theta_{\text{PEE}} - \theta_0| \to_p 0$ and $|\hat \theta_{\text{PDRE}} - \theta_0| \to_p 0$.
    \item Under Assumptions \ref{assump:latentcar}, \ref{assump:proximalpos}, \ref{assump:censoringbridge}, \ref{assump:censoruntestcomplete}, and \ref{assump:uniformcons1}(b) with $q_* = q_0$, $|\hat \theta_{\text{PCE}} - \theta_0| \to_p 0$ and $|\hat \theta_{\text{PDRE}} - \theta_0| \to_p 0$.
\end{enumerate}
\end{thm}

\begin{thm}[Asymptotic Normality]\label{thm:an}
~~~
\begin{enumerate}
    \item Under Assumptions \ref{assump:latentcar}, \ref{assump:proximalpos}, \ref{assump:eventbridge}, \ref{assump:eventuntestcomplete}, and \ref{assump:if}(a) with $h_* = h_0$, both $\sqrt{n}(\hat \theta_{\text{PEE}} - \theta_0)$ and $\sqrt{n}(\hat \theta_{\text{PDRE}} - \theta_0)$ tend to zero mean normal distributions weakly;
    \item Under Assumptions \ref{assump:latentcar}, \ref{assump:proximalpos}, \ref{assump:censoringbridge}, \ref{assump:censoruntestcomplete}, and \ref{assump:if}(b) with $q_* = q_0$, both $\sqrt{n}(\hat \theta_{\text{PCE}} - \theta_0)$ and $\sqrt{n}(\hat \theta_{\text{PDRE}} - \theta_0)$ tend to zero mean normal distributions weakly.  
\end{enumerate}
\end{thm}
The variances of the asymptotic normal distributions depend heavily on the form of models on $h$ and $q$. Since all estimators under the assumptions imposed are regular and asymptotically linear, bootstrap can be used for inference and is applied in the simulation below. \textcolor{black}{We use the random weighting bootstrap method \citep[Chapter 10]{shao2012jackknife}. Take nonparametrically bootstrapping the asymptotic distribution of an $\hat \theta_{\text{PEE}}$ as an example. Suppose we conduct $B$ rounds of bootstrap, for each bootstrap replication $1 \leq b \leq B$, we generate i.i.d. standard exponential random variables $e_{i, b}$ independent of our data. Then we standardize them to get the bootstrap weight $w_{i, b} = ne_{i, b}/\sum_{i = 1}^ne_{i, b}$. We multiply $w_{i, b}$ in \eqref{eq:hmarginsample} to get bootstrap estimate $\hat B_b(t)$. We then plug it in to get bootstrap estimate $\hat H_{i, b}(t; D, \theta)$. Next, our bootstrap estimate $\hat \theta_b$ is derived by solving $\frac{1}{n}\sum_{i = 1}^nw_{i, b}\hat H_{i, b}(t; D, \theta) = o_p(n^{-\frac12})$. Finally, one can use variance of $\hat \theta_b$, $1 \leq b \leq B$, as an estimate of $\Var(\hat \theta)$. The random weighting bootstrap method is more preferable than resampling-based bootstrap for survival analysis, since resampling creates too many ties for survival data. The random weighting bootstrap method is also known as bootstrap clone method \citep{lo1991bayesian}, multiplier bootstrap \citep{kosorok2008introduction} and is closely related to the Bayesian bootstrap \citep{rubin1981bayesian}.}

\section{Empirical Illustrations}\label{sec:example}

\subsection{Concrete Estimators}\label{sec:concrete}
In this section, we illustrate our estimation framework with a concrete example, where we will construct estimators for $h_0$ and $q_0$ such that assumptions in Section \ref{sec:est} hold, and investigate their finite-sample performance. We work under a simple yet heuristic situation when there are only baseline covariates. Importantly, this simplification is without loss of generality as it still captures the essential complexities of the continuous-time dependent right censoring nature. In this vein, since most of our framework is conditioned on the covariates, we may focus on illustration rather than having to consider the complicated structure of the distributions of time-varying covariates. We remark a nonparametric identification for dependent censoring (including competing risks) for studies only involving baseline covariates is still novel in the literature. Suppose we observe $n$ i.i.d. samples $(\widetilde T_i, \Delta_i, X_i, W_i, Z_i)$. We are interested in 
\begin{equation}
    D(T, X; \theta) = \mathbbm{1}(T > 0.5) - \theta_0,
\end{equation}
or equivalently $\theta_0 = \P(T > 0.5)$. In this case $T_D = \min(T, 0.5)$. We will use the estimators based on $(\widetilde T_D, \Delta_D)$, see Remark \ref{rem:td2}. 
We may impose semiparametric models on $h$ and $q$ in the additive form as
\begin{align}\label{eq:simuh}
    &H\{t; D, \theta, B(t)\} = h\{t, W, X; D, \theta, B(t)\} \\
    &=
    \begin{cases}
        \exp\{B_0(t) + B_w(t) W + B_x(t) X\} - \theta,&t \leq 0.5,\\
        1 - \theta, &t > 0.5,
    \end{cases}
\end{align}
and
\begin{equation}\label{eq:simuq}
    Q\{t; A(t)\} = q\{t, Z, X; A(t)\} = \exp\{A_0(t) + A_z(t)Z + A_x(t)X\}.
\end{equation}
These models allow time-varying parameters $B(t)$ and $A(t)$ unspecified, together with the double robustness for $\theta_0$ given in Theorem \ref{thm:an}, alleviating concerns on possible model mis-specification. Note that when ``conditional independent censoring'' \eqref{eq:car} holds given $L(t) = (X(t), W(t))$ such that $Z(t) = W(t)$, \eqref{eq:simuh} and \eqref{eq:simuq} indeed correspond to the additive hazard model \citep[Chapter 5.1]{martinussen2006dynamic}. $B(t)$ and $A(t)$ correspond to the cumulative regression coefficients therein. By \eqref{eq:simuh}, \eqref{eq:simuq}, and the composition function rule for differentiation, the Equations \eqref{eq:hmarginsample} and \eqref{eq:qmarginsample} lead to differential form estimating equations
\begin{align}
    &\sum_{i = 1}^n \mathbbm{1}(\widetilde T_i \geq t)\exp\{\hat B_0(t+) + \hat B_w(t+)W_i + \hat B_x(t+)X_i\}\\
    &~~~~~~(1, Z_i, X_i)^\top \{dB_0(t) + dB_w(t)W_i + dB_x(t)X_i - dN_{T, i}(t)\} = o_p(n^{-\frac{1}{2}}),
\end{align}
and
\begin{align}
    &\sum_{i = 1}^n \mathbbm{1}(\widetilde T_i \geq t)\exp\{\hat A_0(t-) + \hat A_z(t-)Z_i + \hat A_x(t-)X_i\}\\
    &~~~~~~(1, W_i, X_i)^\top \{dA_0(t) + dA_z(t)Z_i + dA_x(t)X_i - dN_{C, i}(t)\} = o_p(n^{-\frac{1}{2}}),
\end{align}
which imply that $\hat B(t)$ and $\hat A(t)$ can be computed recursively, as in \citet{martinussen2006dynamic, martinussen2017instrumental, ying2022structural, ying2022structural2}. Also, $\hat B(t)$ and $\hat A(t)$ only changes at $dN_{T, i}(t) = 1$ and $dN_{C, i}(t) = 1$. Since $D(T, X; \theta) = \mathbbm{1}(T > 0.5) - \theta$, we can set initial values $\hat B(0.5) = \hat A(0) = 0$ to satisfy initial conditions \eqref{eq:hmargininitsample} and \eqref{eq:qmargininitsample}. Therefore we have
\begin{equation}
    \hat B(t) = 
    \begin{cases}
    \frac{1}{n}\int_0^t \{\hat \bbM_B(s)\}^\dagger \sum_{i = 1}^n \mathbbm{1}(\widetilde T_i \geq s)\exp\{\hat B_0(s+) + \hat B_w(s+)W_i + \hat B_x(s+)X_i\}\\
    ~~~~~~~~~~~~~~\cdot (1, Z_i, X_i)^\top dN_i(s), &t \leq 0.5,\\
    0, & t> 0.5,
    \end{cases}
\end{equation}
where $\dagger$ represents the Moore-Penrose inverse,
\begin{equation}
    \hat \bbM_B(t) = \frac{1}{n}\sum_{i = 1}^n \mathbbm{1}(\widetilde T_i \geq t)\exp\{\hat B_0(t+) + \hat B_w(t+)W_i + \hat B_x(t+)X_i\}(1, Z_i, X_i)^\top (1, W_i, X_i),
\end{equation}
and
\begin{equation}
    \hat A(t) = \frac{1}{n}\int_0^t \{\hat \bbM_A(s)\}^\dagger \sum_{i = 1}^n \mathbbm{1}(\widetilde T_i \geq s)\exp\{\hat A_0(s-) + \hat A_z(s-)Z_i + \hat A_x(s-)X_i\}(1, W_i, X_i)^\top dN_i(s),
\end{equation}
where
\begin{equation}
    \hat \bbM_A(t) = \frac{1}{n}\sum_{i = 1}^n \mathbbm{1}(\widetilde T_i \geq t)\exp\{\hat A_0(t-) + \hat A_z(t-)Z_i + \hat A_x(t-)X_i\}(1, W_i, X_i)^\top (1, Z_i, X_i).
\end{equation}
By this way, it is easy to see that both $\hat B(t)$ and $\hat A(t)$ are asymptotically linear and therefore asymptotically normal with minor regularity conditions, like estimators for the cumulative regression coefficients \citep{martinussen2006dynamic}. We write
\begin{equation}
    \hat H_i(t; D, \theta) = h\{t, W_i, X_i; D, \theta, \hat B(t)\},
\end{equation}
and
\begin{equation}
    \hat Q_i(t) = q\{t, W_i, X_i; \hat A(t)\}.
\end{equation}
Therefore, our proximal estimators are computed as
\begin{equation}
    \hat \theta_{\text{PEE}} = \frac{1}{n}\sum_{i = 1}^n \hat H_i(0; D, 0) = \frac{1}{n}\sum_{i = 1}^n \exp\{\hat B_0(0) + \hat B_w(0)W_i + \hat B_x(0)X_i\},
\end{equation}
\begin{align}
    \hat \theta_{\text{PCE}} &= \frac{\sum_{i = 1}^n \Delta_{D, i}\hat Q_i(\widetilde T_{D, i})\mathbbm{1}(\widetilde T_i > 0.5)}{\sum_{i = 1}^n \Delta_{D, i}\hat Q_i(\widetilde T_{D, i})} \\
    &= \frac{\sum_{i = 1}^n \Delta_{D, i}\exp\{\hat A_0(\widetilde T_{D, i}) + \hat A_w(\widetilde T_{D, i})W_i + \hat A_x(\widetilde T_{D, i})X_i\}\mathbbm{1}(\widetilde T_i > 0.5)}{\sum_{i = 1}^n \Delta_{D, i}\exp\{\hat A_0(\widetilde T_{D, i}) + \hat A_w(\widetilde T_{D, i})W_i + \hat A_x(\widetilde T_{D, i})X_i\}},
\end{align}
\begin{align}
    \hat \theta_{\text{PDRE}} 
    &= \frac{\sum_{i = 1}^n \Delta_{D, i} \hat Q_i(\widetilde T_{D, i}) \mathbbm{1}(\widetilde T_i > 0.5) - \sum_{i = 1}^n\int_0^{\widetilde T_{D, i}}\hat H_i(t; D, 0)\{d\hat Q_i(t) - \hat Q_i(t)dN_{C, i}(t)\}}{\sum_{i = 1}^n \Delta_{D, i} \hat Q_i(\widetilde T_{D, i}) - \sum_{i = 1}^n\int_0^{\widetilde T_{D, i}}\{d\hat Q_i(t) - \hat Q_i(t)dN_{C, i}(t)\}}.
\end{align}
\textcolor{black}{Based on the semiparametric form of $H$ and $Q$, our estimators recursively solve linear problems, which can be a lot more convenient computationally then conventional proximal causal inference estimation procedure, which requires calling a solver in R, based on the author's experience \citep{tchetgen2020introduction, cui2020semiparametric, ying2021proximal}.}

\subsection{Simulations}\label{sec:simu}
In this section, we investigate the finite-sample performance of PEE, PCE, and PDRE estimators. We generate data $(X, U, Z, W, C, T)$ as followed:
\begin{equation}
    X \sim \max\{\cN(0.6, 0.45^2), 0\}, 
\end{equation}
\begin{equation}
    U \sim \max\{\cN(0.6, 0.45^2), 0\}, 
\end{equation}
\begin{equation}
    Z \sim \cN(1.4 + 0.3 \cdot X - 0.9 \cdot U, 0.25^2),
\end{equation}
\begin{equation}
    W \sim \cN(0.6 - 0.2 \cdot X + 0.5 \cdot U, 0.25^2),
\end{equation}
\begin{equation}
    \P(T > t|X, U) = \exp\{-(0.25 + 0.3 \cdot X + 0.6 \cdot U)t\},
\end{equation}
\begin{equation}
    \P(C > t|X, U) = 
    \begin{cases}
    \exp(-(0.1 + 0.25 \cdot X + 1 \cdot U)t), & t < 3;\\
    0, &t \geq 3.
    \end{cases}
\end{equation}
Note that this setting fits into the scenario we considered in Section \ref{sec:compatible}. We further verify in Appendix Section \ref{sec:compatibility} that this data generating mechanism is compatible with all Assumptions \ref{assump:latentcar}-\ref{assump:censoruntestcomplete} required for identification and also, the semiparametric models \eqref{eq:simuh} and \eqref{eq:simuq} contain the truth.

Our estimand is
\begin{equation}\label{eq:simumsmm}
    \theta = \P(T > 0.5),
\end{equation}
as the solution to the full data estimating equation $D(T, X; \theta) = \mathbbm{1}(T > 0.5) - \theta$. Each simulated dataset is analyzed using $\widehat \theta_{\text{PEE}}$, $\widehat \theta_{\text{PCE}}$, and $\widehat \theta_{\text{PDRE}}$ given in Section \ref{sec:concrete}. \textcolor{black}{As a comparison, we additionally compute $\widehat \theta_{\text{DRE}}$, which are classical doubly robust estimators assuming conditional independent censoring \eqref{eq:car}, by taking $(X, W, Z)$ as covariates, the Kaplan-meier estimator $\widehat \theta_{\text{KM}}$ that assumes non-informative censoring, and the doubly robust estimator $\widehat \theta_{\text{DRE-truth}}$ that take $(X, U)$ as covariates. Therefore, $\widehat \theta_{\text{PEE}}$, $\widehat \theta_{\text{PCE}}$, $\widehat \theta_{\text{PDRE}}$, and $\widehat \theta_{\text{DRE-truth}}$ are expected to perform well. $\widehat \theta_{\text{DRE-truth}}$ serves as the benchmark because it takes the unobserved true risk factor $U$ as covariates. On the other hand, $\widehat \theta_{\text{DRE}}$ and $\widehat \theta_{\text{KM}}$ are expected to entail bias.}

We examine the performance of these estimators by reporting biases, empirical standard errors (SEE), average estimated standard errors (SD) based on $B = 20$ random weighting bootstrap, and coverage probabilities (95\% CP) of Wald type 95\% confidence intervals using $R = 1000$ simulated data sets of size $n = 1500$. Here we have verified that $B = 20$ is enough for random weighting bootstrap variance estimates to perform well in this case, as indeed shown in the result. \citet[Chapter 6]{efron1994introduction} also mentioned that bootstrap with even a small number of replications can usually estimate variance accurately. The results are given in Table \ref{tab:simubeta500}.

\begin{table}
\caption{\label{tab:simubeta500}
Simulation results of PEE, PCE, and PDRE estimators. We report bias ($\times 10^{-3}$), empirical standard error (SEE) ($\times 10^{-3}$), average estimated standard error (SD) ($\times 10^{-3}$) by $B = 20$ random weighting bootstrap, and coverage probability of Wald type 95\% confidence intervals (95\% CP) of $\widehat \theta_{\text{PEE}}$, $\widehat \theta_{\text{PCE}}$, and $\widehat \theta_{\text{PDRE}}$, for $n = 1500, 3000 $ sample sizes and $R = 1000$ Monte Carlo samples.
}
\centering
\begin{tabular}{|lcccccccc|}
  \hline
  &\multicolumn{2}{c}{n = 1500} && &\multicolumn{2}{c}{n = 3000}&&\\
 & Bias & SEE & SD & 95\% CP & Bias & SEE & SD & 95\% CP \\ 
  \hline
$\widehat \theta_{\text{PEE}}$ & 0.9 & 14.3 & 13.8 & 93.5    & 0.5 & 9.6 & 9.8 & 94.7 \\ 
  $\widehat \theta_{\text{PCE}}$ & 0.7 & 14.4 & 14.0 & 93.6  & 0.5 & 9.6 & 9.8 & 95.0\\ 
  $\widehat \theta_{\text{PDRE}}$ & 0.9 & 14.5 & 14.3 & 93.4 & 0.8 & 9.6 & 9.9 & 95.0 \\ 
  $\widehat \theta_{\text{DRE}}$ & 4.1 & 14.0 & 13.5 & 92.6 & 3.8 & 9.4 & 9.5 & 92.1\\ 
  $\widehat \theta_{\text{KM}}$ & 9.0 & 13.8 & 13.4 & 87.3  & 9.4 & 9.3 & 9.4 & 81.2 \\ 
  $\widehat \theta_{\text{DRE-truth}}$ & 0.80 & 13.7 & 93.3 & 95.5 & -0.1 & 9.5 & 9.7 & 94.4 \\ 
   \hline
\end{tabular}
\end{table}
As the simulation results illustrate, $\widehat \theta_{\text{PEE}}$, $\widehat \theta_{\text{PCE}}$, $\widehat \theta_{\text{PDRE}}$, and $\widehat \theta_{\text{DRE-truth}}$ perform well with small biases, thus confirming our theoretical results. As expected from theory, variance estimates approach the Monte Carlo variance. Wald type confidence intervals of $\widehat \theta_{\text{PEE}}$, $\widehat \theta_{\text{PCE}}$, $\widehat \theta_{\text{PDRE}}$, and $\widehat \theta_{\text{DRE-truth}}$ attain their nominal levels as sample sizes become larger. Both $\widehat \theta_{\text{DRE}}$ and $\widehat \theta_{\text{KM}}$ have substantial biases, compared to other estimators.




\section{\textcolor{black}{A Real Data Application on the SEER-Medicare Dataset}}\label{sec:real}
The publicly available SEER-Medicare dataset presents a unique opportunity to delve into the complex interplay between health outcomes and the human experience. This rich tapestry of information, woven from cancer diagnoses, healthcare utilization, and demographic details, empowers researchers to explore a multitude of questions crucial for public health. In this real-data application, we harness the power of SEER-Medicare to tackle the intertwined issues of all-cause mortality and loss to follow-up (LTFU).


Understanding the interplay between loss to follow-up (LTFU) and all-cause mortality is crucial. These seemingly separate phenomena are often intimately connected, and ignoring this link can lead to biased and misleading conclusions if we apply simple statistical methods assuming independence. Specific factors like socioeconomic status and treatment strategies play a role in this intricate interplay. Lower socioeconomic status can limit access to healthcare and transportation, hindering follow-up participation. Additionally, it can be associated with poorer overall health and higher mortality rates. This creates a tangled web where both LTFU and mortality are influenced by the same factors, causing biased estimates if treated separately. On the other hand, aggressive treatment regimens associated with certain cancers may deter participation due to physical or emotional toll, while simultaneously impacting overall survival. This again creates a dependence between LTFU and mortality, biasing analyses that assume otherwise. Assuming independence when these factors are intertwined can lead to a variety of biases. Failing to account for informative LTFU might lead to underestimating mortality rates, particularly for high-risk groups who are more likely to drop out. To avoid these pitfalls, researchers must use statistical methods that acknowledge the dependence between LTFU and mortality. These methods, such as survival analysis with proper handling of censoring mechanisms, can provide more accurate and reliable estimates.

We apply our method onto a randomly sampled subset of the SEER-Medicare dataset with sample size $n = 10000$. We define $T$ as time to death, $C$ as time of loss to follow up. We consider covariates $X$ to include age, race, and sex. We set $Z$ as median household income and rural urban continuum code. We set $W$ as treatment strategy and tumor grade. As we mentioned earlier, socioeconomic status and treatment strategies play an important role in causing dependency structure between LTFU and all cause mortality. Household income and rural urban continuum code are proxies of underlying unknown socioeconomic status, might influence patient's decision of participation of the study but this recorded level is not a direct cause of all cause mortality. On the other hand, treatment strategy might influence all cause mortality but is a proxy of underlying unknown physical or emotional toll that deters participation. Tumor grade is a proxy of prognosis which might influence participation and also predict mortality but itself it not a direct cause.

The SEER-Medicare dataset has been tracking cancer incidence and survival data for Medicare beneficiaries since 1991, with the first linkage of SEER data and Medicare claims occurring in 1992. The dataset is updated periodically with new data, with the most recent update including data through 2020. Therefore, with 30 years of data, we decide to use our framework to estimate $\P(T > 5)$, $\P(T > 15)$, and $\P(T > 25)$, respectively. As benchmarks, we also apply the Kaplan-Meier estimator assuming non-informative and AIPCW estimator using $L = (X, Z, W)$ as covariates. Table \ref{tab:real} shows the results.

When $T > 5$, PEE gives a smaller estimate than PCE, PDRE, DRE, and KM, which are closer to each other. As time goes larger when $T > 15$ and $T > 25$, all proximal estimators (PEE, PCE, PDRE) give smaller estimates than that of DRE and KM. This suggests that proximal estimators might suggest possible existence of latent factors leading to informative censoring, which the Kaplan-Meier estimator may overestimate due to its inability to accommodate such complexities. Although the variance estimates for PEE, PCE, and PDRE are notably broader, this may stem from the inherent challenges associated with the Fredholm integral equation or indicate the need for more refined variance estimation techniques. 

Our approach, breaking new ground in survival analysis, introduces a novel proximal survival analysis framework. This method does not rely on stringent assumptions like `conditional independent censoring', which is often unrealistic in complex observational data. Our work represents a significant departure from traditional methods, leveraging proxy variables and a more realistic acknowledgment of unmeasured covariates. Despite the larger estimated variances, our framework's theoretical contributions and the flexibility it offers in handling unmeasured confounding in messy observational data underscore its potential as a valuable alternative in survival analysis.
\begin{table}
\caption{\label{tab:real}
Results of real data application of the SEER-Medicare dataset. We report point estimates from PEE, PCE, PDRE, standard DRE, and KM estimator, together with their 95\% confidence intervals (multiplier bootstrap with $B = 20$) in parentheses.
}
\centering
\begin{tabular}{|lccc|}
  \hline
 & $\hat \P(T > 5)$ & $\hat \P(T > 15)$ & $\hat \P(T > 25)$  \\ 
  \hline
PEE & 0.526 (0.330, 0.722) & 0.257 (0.013, 0.501) & 0.108 (0.000, 0.310)  \\ 
  PCE & 0.673 (0.501, 0.846) & 0.183 (0.000, 0.578) & 0.021 (0.000, 0.092)  \\ 
  PDRE & 0.686 (0.517, 0.856) & 0.263 (0.000, 0.574) & 0.101 (0.000, 0.243) \\ 
  DRE & 0.642 (0.633, 0.651)  & 0.372 (0.360, 0.388) & 0.186 (0.173, 0.199) \\ 
  KM & 0.642 (0.633, 0.651) & 0.375 (0.361, 0.389) & 0.191 (0.176, 0.205)\\ 
   \hline
\end{tabular}
\end{table}

\section{Discussion}\label{sec:dis}
In this paper, we caution practitioners to think twice about the validity of ``conditional independent censoring'' and encourage them to consider our proximal framework to deal with dependent censoring, which cannot be explained away by observed covariates, by formally admitting that measured covariates are at best imperfect proxies of underlying censoring mechanism. The framework essentially hinges on an analyst's ability to classify the observed covariates processes into three types, with sufficient completeness assumptions, without any stringent restrictions on the observed data distribution. We present two approaches for identification that later yield a doubly robust identification that is more robust against possible model mis-specification. We construct estimators based on semiparametric models with asymptotic inferential tools. For illustration, we consider a simple yet heuristic enough setting where we only have baseline covariates but informative censoring that cannot be explained away by these covariates. We examine the finite-sample performance of these estimators via a Monte-Carlo simulation, with a real data application on the SEER-Medicare dataset. One can also check how close results are, with and without ``conditional independent censoring'', as a way to examine the robustness of their conclusion. 

As we mentioned in the introduction, our framework borrows ideas from ``proximal causal inference''. We remark that ``proximal causal inference'' has only been most developed in longitudinal studies when time advances in discrete steps for causal inference. Moreover, empirical solutions to those integral equations in Assumptions \ref{assump:eventbridge} and \ref{assump:censoringbridge} are notoriously challenging to compute due to the ill-posedness nature of the problem \citep{ai2003efficient}. Our current survival analysis setting is considerably more challenging than considered in prior works, as the time to event, censoring time, and time-varying covariates are allowed to be measured in continuous time. Therefore our development of identification and estimation to survival analysis is highly non-trivial, both theory-wise and computation-wise.

There are several possible future directions for this line of research. First of all, nonparametric estimation like \citet{ghassami2022minimax} for nuisance functions $(H(t; D, \theta), Q(t))$ can be developed. \textcolor{black}{However, one should note that Fredholm integral equations that are central to our identification are ill-posed in the sense that the solution varies discontinuously with the initial condition. This presents a major complication for nonparametric estimation and inference based on this kind of integral equations.} Semiparametric theory can be considered, which can provide more understanding of the efficiency of possible estimators. The extension to a more general coarsened data structure and also biased sampling data can be considered. The study of the existence of a process to the integral equations in the form of \eqref{eq:eventconfbridgeiden}, \eqref{eq:eventconfbridgeideninit} or \eqref{eq:censorconfbridgeiden}, \eqref{eq:censorconfbridgeideninit} is undoubtedly of probabilistic interest.

\section*{Acknowledgements}
The author would like to thank Eric J. Tchetgen Tchetgen, Ronghui (Lily) Xu for valuable suggestions for improving the manuscript.

\appendix
\section*{Description of the Supplementary Material}
The supplementary material includes proofs to all results given in the main text of the paper. We also provide scenarios under which Assumptions \ref{assump:eventbridge}, \ref{assump:censoringbridge} hold. We verify that our simulation setting satisfies Assumption \ref{assump:latentcar}-\ref{assump:censoruntestcomplete}. We verify that our concrete estimators satisfy Assumption \ref{assump:uniformcons1}, \ref{assump:if}.

Code replicating numerical results including all simulations are provided on github. Here is the link: \url{https://github.com/andrewyyp/Proximal_Survival_Analysis}. The SEER-Medicare dataset we have used in the real data application is publicly available through \url{https://healthcaredelivery.cancer.gov/seermedicare/obtain/}.

\bibliographystyle{agsm}

\bibliography{ref, lpci_ref}

\newpage

\section{\textcolor{black}{Plausibility of Assumptions \ref{assump:eventbridge}, \ref{assump:censoringbridge}}}\label{sec:compatible}
We understand that the Assumptions \ref{assump:eventbridge}, \ref{assump:censoringbridge} may be strong assumptions. The corresponding existence condition in standard proximal causal inference is already rather unintuitive but there are at least available sufficient conditions in the form of Picard's criterion plus an additional statistical completeness condition. Moreover, in the standard setting there are widely-used parametric models (e.g., the multivariate Gaussian model) in which a bridge function can be shown to exist. The existence of the bridge process is central to our analysis, and in fact the estimation results require not only that at least one of two bridge processes exists but that it has a particular parametric functional form. Here we want to at least provide a parametric example where such a bridge process can be shown to exist to assure our readers. Note that the following example is indeed quite general and our framework is considered to be more general than this.

We consider a data generating process where $(\overline U(t), \overline X(t))$ is an arbitrary multivariate Gaussian process, $W(t)$, $Z(t)$ are generated from some linear combination of $(U(t), X(t))$ plus some Gaussian noise, and $T$, $C$ follow
\begin{equation}
    \lambda_T\{t|\overline U(\infty), \overline X(\infty), \overline W(\infty), \overline Z(\infty)\} = \beta_0(t) + \beta_x(t) X(t) + \beta_u(t) U(t),
\end{equation}
\begin{equation}
    \lambda_C\{t|\overline U(\infty), \overline X(\infty), \overline W(\infty), \overline Z(\infty)\} = \alpha_0(t) + \alpha_x(t) X(t) + \alpha_u(t) U(t),
\end{equation}
for some $(\beta_0(t), \beta_x(t), \beta_u(t))$ and $(\alpha_0(t), \alpha_x(t), \alpha_u(t))$. Suppose we are interested in
\begin{equation}
    \theta = \P(T > \tau),
\end{equation}
for some constant $\tau > 0$. Consider
\begin{align}
    &H\{t; D, \theta, B(t)\} = h\{t, \overline W(t), \overline X(t); D, \theta, B(t)\} \\
    &=
    \begin{cases}
        \exp\{B_0(t) + B_x(t) X(t) + B_w(t) W(t)\} - \theta,  &t \leq \tau,\\
        1 - \theta, &t > \tau.
    \end{cases}
\end{align}

Clearly \eqref{eq:eventbridgeUinit} is satisfied. Then one can show that
\begin{align}
    &\E[dH\{t; D, \theta, B(t)\} - H\{t-; D, \theta, B(t-)\}dN_T(t)|\widetilde T \geq t, \overline Z(t-), \overline X(t)\}\\
    &=\E\{\E[dH\{t; D, \theta, B(t)\} - H\{t-; D, \theta, B(t-)\}dN_T(t)\\
    &~~~~|\widetilde T \geq t, \overline U(t), \overline X(t), \overline W(t-), \overline Z(t-)\}|\widetilde T \geq t, \overline X(t), \overline Z(t-)\}]\\
    &=\E(H\{t-; D, \theta, B(t-)\}[\E\{dB_0(t) + dB_x(t) X(t) + dB_w(t) W(t) \\
    &~~~~- dN_T(t)|\widetilde T \geq t, \overline U(t), \overline X(t), \overline W(t-), \overline Z(t-)\}|\widetilde T \geq t, \overline X(t), \overline Z(t-)\}]\\
    &=\E(H\{t-; D, \theta, B(t-)\}\\
    &~~~[\E\{dB_0(t) + dB_x(t) X(t) + dB_w(t) W(t)|\widetilde T \geq t, \overline U(t), \overline X(t), \overline W(t-), \overline Z(t-)\} \\
    &~~~~- \lambda_T\{t|\overline U(t), \overline X(t), \overline W(t-), \overline Z(t-)\}dt|\widetilde T \geq t, \overline U(t-), \overline X(t-)\}].
\end{align}
Therefore to prove $H\{t; D, \theta, B(t)\}$ satisfies \eqref{eq:eventbridgeU}, it suffices to find such that
\begin{align}
    &\E\{dB_0(t) + dB_x(t) X(t) + dB_w(t) W(t)|\widetilde T \geq t, \bar W(t-), \bar Z(t-), \bar X(t), \bar U(t)\} \\
    &=\lambda_T\{t|\overline U(t), \overline X(t), \overline W(t-), \overline Z(t-)\}dt\\
    &= \beta_0(t)dt + \beta_x(t) X(t)dt + \beta_u(t) U(t)dt.\label{eq:solution}
\end{align}
Note that $W(t)$ and $Z(t)$ are linear combination of $X(t), U(t)$ plus some Gaussian noise, and $T$ and $C$ only depend on $\bar X(t)$, $\bar U(t)$, therefore
\begin{align}
    &\E\{B_0(t) + B_x(t) X(t) + B_w(t) W(t)|\widetilde T \geq t, \bar W(t), \bar Z(t), \bar X(t), \bar U(t)\} \\
    &= \E\{B_0(t) + B_x(t) X(t) + B_w(t) W(t)|X(t), U(t)\}\\
    &= B_0(t) + B_x(t) X(t) + B_w(t)\E\{W(t)|X(t), U(t)\}.
\end{align}
Note that $\E\{W(t)|X(t), U(t)\}$ is linear in $X(t), U(t)$, therefore one is able to find $B(t)$ such that \eqref{eq:solution} holds. Consider
\begin{equation}
    Q\{t; A(t)\} = \exp\left[\int_0^t\left\{A_0(s) + A_X(s) X(s) + A_Z(s) Z(s)\right\}ds\right].
\end{equation}
Clearly all $Q(t)$ satisfies \eqref{eq:censorbridgeUinit}. The proof of $Q\{t; A(t)\}$ satisfying \eqref{eq:censorbridgeU} follows similarly as that of $H\{t; D, \theta, B(t)\}$.

\section{Proofs}\label{sec:proofs}
\subsection{Proof of Theorem \ref{thm:eventiden}}
The proof will process as follows: firstly, we show that the observed integral equation \eqref{eq:eventconfbridgeiden} implies its latent counterpart \eqref{eq:eventbridgeU} through Assumptions \ref{assump:latentcar} and \ref{assump:eventuntestcomplete}. Then the $h(t)$ function plays a role as the project of $D\{T, \overline X(T); \theta\}$ onto latent factors up to time $t$.

By an iterative application of the Assumption \ref{assump:latentcar}, it is immediate to show
\begin{align}
    &\E\left[\E\left\{dH(t; D, \theta) - H(t; D, \theta)dN_T(t)|T \geq t, \overline U(t), \overline X(t)\right\}|\widetilde T \geq t, \overline Z(t), \overline X(t)\right]\\
    &=\E\left[\E\left\{dH(t; D, \theta) - H(t; D, \theta)dN_T(t)|T \geq t, C \geq t, \overline Z(t), \overline U(t), \overline X(t)\right\}|\widetilde T \geq t, \overline Z(t), \overline X(t)\right]\\
    &=\E\left[dH(t; D, \theta) - H(t; D, \theta)dN_T(t)|\widetilde T \geq t, \overline Z(t), \overline X(t)\right] = 0,
\end{align}
and
\begin{align}
    &\E\left(\E\left[D\{T, \overline X(T); \theta\} - H(T; D, \theta)|\overline U(T), \overline X(T)\right]|\Delta = 1, \overline Z(T), \overline X(T)\right)\\
    &\E\left(\E\left[D\{T, \overline X(T); \theta\} - H(T; D, \theta)|T \geq T, \overline U(T), \overline X(T)\right]|\Delta = 1, \overline Z(T), \overline X(T)\right)\\
    &=\E\Big(\E\left[D\{T, \overline X(T); \theta\} - H(T; D, \theta)|C \geq T, T \geq T, \overline U(T), \overline Z(T), \overline X(T)\right]\\
    &~~~~~~~|\Delta = 1, \overline Z(T), \overline X(T)\Big)\\
    &=\E\left\{D\{T, \overline X(T); \theta\} - H(T; D, \theta)|\Delta = 1, \overline Z(T), \overline X(T)\right\} = 0,
\end{align}
which by Assumption \ref{assump:eventuntestcomplete} yields \eqref{eq:eventbridgeU} and \eqref{eq:eventbridgeUinit}. Indeed, with integration by parts we have
\begin{align}
    Q(T)H(T; D, \theta)- Q(0)H(0; D, \theta) &= \int_0^{T} Q(t)\{dH(t; D, \theta) - H(t; D, \theta)dN_T(t)\}\\
    &~~+ \int_0^{T} H(t; D, \theta)dQ(t),
\end{align}
for any function $Q(t) = q\{t, \overline U(t), \overline X(t)\}$. In particular, setting $Q(t) \equiv 1$ and taking expectation we have
\begin{align}
    &\E\{D(T, \overline X(T); \theta)\} - \E\{H(0; D, \theta)\} \\
    &= \E\{Q(T)H(T; D, \theta)\} - \E\{Q(0)H(0; D, \theta)\} \\
    &= \E\left[\int_0^{T} Q(t)\{dH(t; D, \theta) - H(t; D, \theta)dN_T(t)\}\right] + \E\left\{\int_0^{T} H(t; D, \theta)dQ(t)\right\}\\
    &= 0.
\end{align}

\subsection{Proof of Theorem \ref{thm:censoringiden}}
The proof will process similar as in the proof of Theorem \ref{thm:eventiden} as follows: firstly, we show that the observed integral equation \eqref{eq:censorbridgeiden} implies its latent counterpart \eqref{eq:censorbridgeU} through Assumptions \ref{assump:latentcar} and \ref{assump:censoruntestcomplete}. Then the $q(t)$ function will play a role as the latent inverse probability censoring weight to eliminate bias introduced by censoring.

It is immediate to show that
\begin{align}
    &\E\left[\E\left\{dQ(t) - Q(t)dN_C(t)|T \geq t, C \geq t, \overline U(t), \overline X(t)\right\}|\widetilde T \geq t, \overline W(t), \overline X(t)\right]\\
    &=\E\left[\E\left\{dQ(t) - Q(t)dN_C(t)|T \geq t, C \geq t, \overline W(t), \overline U(t), \overline X(t)\right\}|\widetilde T \geq t, \overline W(t), \overline X(t)\right]\\
    &=\E\left\{dQ(t) - Q(t)dN_C(t)|\widetilde T \geq t, \overline W(t), \overline X(t)\right\} = 0.
\end{align}
which by Assumption \ref{assump:censoruntestcomplete} yields \eqref{eq:censorbridgeU} by rearranging terms. We show that the expectation of \eqref{eq:censorbridgeiden} matches that of \eqref{eq:fulldataX},
\begin{align}
    &\E\left\{\Delta Q_{T} D\{T, \overline X(T); \theta\}\right\}\\
    &=\E\left[\E\left\{Q_{T}|C \geq T, T, \overline U(T), \overline X(T)\right\}\Delta D\{T, \overline X(T); \theta\}\right]\\
    &=\E\left[\Prodi_0^T\frac{\E\left\{Q(s)|C \geq s + ds, T, \overline U(T), \overline X(T)\right\}}{\E\left\{Q(s - ds)|C \geq s, T, \overline U(T), \overline X(T)\right\}}\Delta D\{T, \overline X(T); \theta\}\right]\\
    &=\E\left[\Prodi_0^T\frac{\E\left\{Q(s)|C \geq s + ds, T \geq s, \overline U(s), \overline X(s)\right\}}{\E\left\{Q(s - ds)|C \geq s, T \geq s, \overline U(s), \overline X(s)\right\}}\Delta D\{T, \overline X(T); \theta\}\right]\\
    &=\E\left[\frac{\Delta D\{T, \overline X(T); \theta\}}{\Prodi_0^T\P\{C \geq s + ds|C \geq s, T \geq s, \overline U(s), \overline X(s))\}}\right]\\
    &=\E\left[\frac{\Delta D\{T, \overline X(T); \theta\}}{\Prodi_0^T\P\{C \geq s + ds|C \geq s, T, \overline U(T), \overline X(T))\}}\right]\\
    &=\E\left[\frac{\Delta D\{T, \overline X(T); \theta\}}{\P\{C \geq T|T, \overline U(T), \overline X(T)\}}\right]\\
    &=\E\left[D\{T, \overline X(T); \theta\}\right].
\end{align}

\subsection{Proof of Theorem \ref{thm:driden}}
When $H(t; D, \theta) = H_0(t; D, \theta)$, by Assumption \ref{assump:eventbridge} and Theorem \ref{thm:eventiden} we have
\begin{align}
    &\E\{\Xi(H_0, Q; D, \theta)\}\\
    &=\E\left[\int_0^{\widetilde T} Q(t)\{dH_0(t; D, \theta) - H_0(t; D, \theta)dN_T(t)\} + H_0\right]\\
    &=\E\left\{H(0; D, \theta)\right\} = \E\left[D\{T, \overline X(T); \theta\}\right].
\end{align}
When $Q(t) = Q_0(t)$, by Assumption \ref{assump:censoringbridge} and Theorem \ref{thm:censoringiden} we have
\begin{align}
    &\E\{\Xi(H, Q_0; D, \theta)\}\\
    &=\E\left[\Delta Q(\widetilde T) D\{\widetilde T, \overline X(\widetilde T); \theta\} - \int_0^{\widetilde T} H(t; D, \theta)\{dQ(t) - Q(t)dN_C(t)\}\right]\\
    &=\E\left[\Delta Q(\widetilde T) D\{\widetilde T, \overline X(\widetilde T); \theta\}\right] =\E\left[D\{T, \overline X(T); \theta\}\right].
\end{align}
\qed

We prepare the following decomposition of the observed data estimating equation, which turns out useful throughout the proofs of both Theorem \ref{thm:cons} and Theorem \ref{thm:an} below,
\begin{equation}\label{eq:decompose}
    \frac{1}{n}\sum_{i = 1}^{n}\Xi_i\{\hat H(t; D, \theta), \hat Q(t)\} = \cT_1 + \cT_2 + \cT_3 + \cT_4 + \cT_5 + \cT_6,
\end{equation}
where
\begin{align}
    \cT_1 &:= \frac{1}{n}\sum_{i = 1}^{n}\Xi_i\{\hat H(t; D, \theta), \hat Q(t)\} - \frac{1}{n}\sum_{i = 1}^{n}\Xi_i\{\hat H(t; D, \theta), Q_*(t)\}\\
    &~~~- \frac{1}{n}\sum_{i = 1}^{n}\Xi_i\{H_*(t; D, \theta), \hat Q(t)\} + \frac{1}{n}\sum_{i = 1}^{n}\Xi_i\{H_*(t; D, \theta), Q_*(t)\}\\
    &= \frac{1}{n} \sum_{i = 1}^n\int_0^\infty \{\hat H_i(t; D, \theta) - H_{*, i}(t; D, \theta)\}d\{\hat Q_i(t) - Q_{*, i}(t)\},\\
    \cT_2 &:= \frac{1}{n}\sum_{i = 1}^{n}\Xi_i\{\hat H(t; D, \theta), Q_*(t)\} - \frac{1}{n}\sum_{i = 1}^{n}\Xi_i\{H_*(t; D, \theta), Q_*(t)\}\\
    &= \frac{1}{n} \sum_{i = 1}^n\int_0^\infty \{\hat H_i(t; D, \theta) - H_{*, i}(t; D, \theta)\}dQ_{*, i}(t),\\
    \cT_3 &:= \frac{1}{n}\sum_{i = 1}^{n}\Xi_i\{H_*(t; D, \theta), \hat Q(t)\} - \frac{1}{n}\sum_{i = 1}^{n}\Xi_i\{H_*(t; D, \theta), Q_*(t)\}\\
    &= \frac{1}{n} \sum_{i = 1}^n\int_0^\infty H_{*, i}(t; D, \theta)d\{\hat Q_i(t) - Q_{*, i}(t)\},\\
    \cT_4 &:= \frac{1}{n}\sum_{i = 1}^{n}\Xi_i\{H_*(t; D, \theta), Q_*(t)\} - \E[\Xi\{H_*(t; D, \theta), Q_*(t)\}]\\
    &~~~- \frac{1}{n}\sum_{i = 1}^{n}\Xi_i\{H_*(t; D, \theta_0), Q_*(t)\} + \E[\Xi\{H_*(t; D, \theta_0), Q_*(t)\}],\\
    \cT_5 &:= \frac{1}{n}\sum_{i = 1}^{n}\Xi_i\{H_*(t; D, \theta_0), Q_*(t)\} - \E[\Xi\{H_*(t; D, \theta_0), Q_*(t)\}],\\
    \cT_6 &:= \E[\Xi\{H_*(t; D, \theta_0), Q_*(t)\}].
\end{align}

\subsection{Proof of Theorem \ref{thm:cons}}
To show the double robustness claimed in Theorem \ref{thm:cons}, it suffices to show that when either $h_* = h_0$ or $q_* = q_0$, the estimating equation $\frac{1}{n}\sum_{i = 1}^{n}\Xi_i\{\hat H(t; D, \theta), \hat Q(t)\} = 0$ remains unbiased. To that end, by \eqref{eq:decompose}, we will prove each term converges to zero one by one. First of all, with Markov's inequality and Holder's inequality we have
\begin{align}
    &\P(|\cT_1| > \delta) \\
    &\leq \frac{1}{\delta}\E\left[\left|\frac{1}{n} \sum_{i = 1}^n\int_0^\infty \{\hat H_i(t; D, \theta) - H_{*, i}(t; D, \theta)\}d\{\hat Q_i(t) - Q_{*, i}(t)\}\right|\right] \\
    &= \frac{1}{\delta}\E\left[\left|\int_0^\infty \{\hat H(t; D, \theta) - H_*(t; D, \theta)\}d\{\hat Q(t) - Q_*(t)\}\right|\right] \\
    &\leq \frac{1}{\delta}\E\left[\int_0^\infty |\hat H(t; D, \theta) - H_0(t; D, \theta)|dV_0^\infty\{\hat Q(t) - Q_*(t)\}\right] \\
    &\leq \frac{1}{\delta}\|\hat H(t; D, \theta) - H_*(t; D, \theta)\|_{\sup, 2}\{\|\hat Q(t)\|_{\text{TV}, 2} + \|Q_*(t)\|_{\text{TV}, 2}\} = o(1),
\end{align}
where the last step results from Assumption \ref{assump:uniformcons1}(a) or (b). Similarly, we have
\begin{align}
    \P(|\cT_2| > \delta) 
    &\leq \frac{1}{\delta}\E\left[\left|\frac{1}{n} \sum_{i = 1}^n\int_0^\infty \{\hat H_i(t; D, \theta) - H_{*, i}(t; D, \theta)\}dQ_{*, i}(t)\right|\right] \\
    &= \frac{1}{\delta}\E\left[\left|\int_0^\infty \{\hat H(t; D, \theta) - H_*(t; D, \theta)\}dQ_*(t)\right|\right] \\
    &\leq \frac{1}{\delta}\E\left[\int_0^\infty |\hat H(t; D, \theta) - H_*(t; D, \theta)|dV_0^\infty\{Q_*(t)\}\right] \\
    &\leq \frac{1}{\delta}\|\hat H(t; D, \theta) - H_*(t; D, \theta)\|_{\sup, 2}\|Q_*(t)\|_{\text{TV}, 2} = o(1),
\end{align}
where the last step again results from Assumption \ref{assump:uniformcons1}(a). For $\cT_3$, we first apply integration by parts to get,
\begin{align}
    \cT_3 &= \frac{1}{n} \sum_{i = 1}^n\int_0^\infty H_{*, i}(t; D, \theta)d\{\hat Q_i(t) - Q_{*, i}(t)\}\\
    &= \frac{1}{n} \sum_{i = 1}^nH_{*, i}(\infty; D, \theta)\{\hat Q_i(\infty) - Q_{*, i}(\infty)\} - \frac{1}{n} \sum_{i = 1}^nH_{*, i}(0; D, \theta)\{\hat Q_i(0) - Q_{*, i}(0)\}\\
    &- \frac{1}{n} \sum_{i = 1}^n\int_0^\infty\{\hat Q_i(t) - Q_{*, i}(t)\} dH_{*, i}(t; D, \theta),
\end{align}
where the second item is zero because of the initial condition $Q(0) \equiv 1$. Next, we have
\begin{align}
    \P(|\cT_3| > \delta) 
    &\leq \frac{1}{\delta}\E\left[\left|\frac{1}{n} \sum_{i = 1}^nH_{*, i}(\infty; D, \theta)\{\hat Q_i(\infty) - Q_{*, i}(\infty)\}\right|\right] \\
    &+ \frac{1}{\delta}\E\left[\left|\frac{1}{n} \sum_{i = 1}^n\int_0^\infty \{\hat Q_i(t) - Q_{0, i}(t)\}dH_{*, i}(t; D, \theta)\right|\right] \\
    &= \frac{1}{\delta}\E\left[\left|H_{*}(\infty; D, \theta)\{\hat Q(\infty) - Q_{*}(\infty)\}\right|\right] \\
    &+ \frac{1}{\delta}\E\left[\left|\frac{1}{n} \sum_{i = 1}^n\int_0^\infty \{\hat Q(t) - Q_{0}(t)\}dH_{*}(t; D, \theta)\right|\right] \\
    &\leq \frac{1}{\delta}\E\left[\sup_t|H_{*}(t; D, \theta)|\sup_t|\hat Q(t) - Q_{*}(\infty)\}|\right] \\
    &+ \frac{1}{\delta}\E\left[\int_0^\infty |\hat Q(t) - Q_*(t)|dV_0^\infty\{H_*(t; D, \theta)\}\right] \\
    &\leq \frac{2}{\delta}\|\hat Q(t) - Q_*(t)\|_{\sup, 2}\|H_*(t; D, \theta)\|_{\text{TV}, 2} = o(1),
\end{align}
by Assumption \ref{assump:uniformcons1}(b). $\cT_4$ by \citet{van1997weak} is $o_p(1)$. $\cT_5$ by the weak law of large numbers converges to 0 in probability. $\cT_6 = 0$ by Theorem \ref{thm:driden} when either $h_* = h_0$ or $q_* = q_0$.

\subsection{Proof of Theorem \ref{thm:an}}
~~~

\noindent
\textsc{Proof of Part}(a):
It suffices to prove part (a) for $\hat \theta_{\text{PDRE}}$ since it is easy to see that $\hat \theta_{\text{PEE}} = \hat \theta_{\text{PDRE}}$ when plugging in $\hat Q = 0 = Q_*$. Now by Assumption \ref{assump:if}(c),
\begin{align*}
    &\sqrt{n}\cT_1 \\
    &= \frac{1}{\sqrt{n}}\sum_{i = 1}^{n}\int_0^{\widetilde T_i} \{\hat H_i(t; D, \beta) - H_0(t; D, \beta)\}\\
    &~~~~\left[d\{\hat Q_i(t) - Q_{*, i}(t)\} - \{\hat Q_i(t) - Q_{*, i}(t)\}dN_{C, i}(t) \right] \\
    &= \frac{1}{\sqrt{n}}\sum_{i = 1}^{n}\int_0^{\widetilde T_i} \frac{1}{n}\sum_{j = 1}^{n}\xi_{h, j}\{t, \overline W_i(t), \overline X_i(t)\}\\
    &~~~\cdot\left[d\frac{1}{n}\sum_{k = 1}^{n}\xi_{q, k}\{t, \overline W_i(t), \overline X_i(t)\} - \frac{1}{n}\sum_{k = 1}^{n}\xi_{q, k}\{t, \overline W_i(t), \overline X_i(t)\}dN_{C, i}(t) \right] +o_p(1),
\end{align*}
which by Chebyshev's inequality we have
\begin{align}
    &\P(\sqrt{n}|\cT_1| > \delta) \\
    &\leq \frac{1}{\delta^2}\Var\Bigg(\frac{1}{\sqrt{n}}\sum_{i = 1}^{n}\int_0^{\widetilde T_i} \frac{1}{n}\sum_{j = 1}^{n}\xi_{h, j}\{t, \overline W_i(t), \overline X_i(t)\}\\
    &~~~\left[d\frac{1}{n}\sum_{k = 1}^{n}\xi_{q, k}\{t, \overline W_i(t), \overline X_i(t)\} - \frac{1}{n}\sum_{k = 1}^{n}\xi_{q, k}\{t, \overline W_i(t), \overline X_i(t)\}dN_{C, i}(t) \right]\Bigg)\\
    &= \frac{1}{\delta^2n^5}\E\Bigg\{\bigg(\sum_{i = 1}^{n}\int_0^{\widetilde T_i} \sum_{j = 1}^{n}\xi_{h, j}\{t, \overline W_i(t), \overline X_i(t)\}\\
    &~~~\left[d\sum_{k = 1}^{n}\xi_{q, k}\{t, \overline W_i(t), \overline X_i(t)\} - \sum_{k = 1}^{n}\xi_{q, k}\{t, \overline W_i(t), \overline X_i(t)\}dN_{C, i}(t) \right]\bigg)^2\Bigg\}\\
    &= \frac{1}{\delta^2n^5}\sum_{1 \leq i, i', j, j', k, k' \leq n}\E\Bigg\{\bigg(\int_0^{\widetilde T_i} \xi_{h, j}\{t, \overline W_i(t), \overline X_i(t)\}\\
    &~~~\left[d\xi_{q, k}\{t, \overline W_i(t), \overline X_i(t)\} - \xi_{q, k}\{t, \overline W_i(t), \overline X_i(t)\}dN_{C, i}(t) \right]\bigg)\\
    &~~~\Bigg(\int_0^{\widetilde T_{i'}} \xi_{h, j'}\{t, \overline W_{i'}(t), \overline X_{i'}(t)\}\\
    &~~~~~~~\left[d\xi_{q, k'}\{t, \overline W_{i'}(t), \overline X_{i'}(t)\} - \xi_{q, k'}\{t, \overline W_{i'}(t), \overline X_{i'}(t)\}dN_{C, i'}(t) \right]\Bigg)\Bigg\}\\
    &= \frac{1}{\delta^2n^5}\Bigg(\sum_{1 \leq i, i', j, j', k, k' \leq n: i, i', j, j', k, k'\text{unequal to each other}} \\
    &~~~~~~~+ \sum_{1 \leq i, i', j, j', k, k' \leq n: i, i', j, j', k, k' \text{at least two are equal}}\Bigg)\\
    &\E\Bigg\{\left(\int_0^{\widetilde T_i} \xi_{h, j}\{t, \overline W_i(t), \overline X_i(t)\}\left[d\xi_{q, k}\{t, \overline W_i(t), \overline X_i(t)\} - \xi_{q, k}\{t, \overline W_i(t), \overline X_i(t)\}dN_{C, i}(t) \right]\right)\\
    &~~~\Bigg(\int_0^{\widetilde T_{i'}} \xi_{h, j'}\{t, \overline W_{i'}(t), \overline X_{i'}(t)\}\\
    &~~~\left[d\xi_{q, k'}\{t, \overline W_{i'}(t), \overline X_{i'}(t)\} - \xi_{q, k'}\{t, \overline W_{i'}(t), \overline X_{i'}(t)\}dN_{C, i'}(t) \right]\Bigg)\Bigg\}.
\end{align}
Note that for the first sum $\sum_{1 \leq i, i', j, j', k, k' \leq n: i, i', j, j', k, k'\text{unequal to each other}}$, all expectations in the sum are zero. The second sum is $O(n^5)$. However, such a bound is not fine enough. We leverage the independence among samples and the fact that $\E[\xi_{h}\{t, \overline w(t), \overline x(t)\}] = \E[\xi_{q}\{t, \overline z(t), \overline x(t)\}] = 0$ to rule out those mean zero terms in the sum $\sum_{1 \leq i, i', j, j', k, k' \leq n}$ to prove that $\sum_{1 \leq i, i', j, j', k, k' \leq n: i, i', j, j', k, k' \text{at least two are equal}}$ is indeed $O(n^4)$. We enumerate all ${6 \choose 2} = 15$ possible cases in $\sum_{1 \leq i, i', j, j', k, k' \leq n: i, i', j, j', k, k' \text{at least two are equal}}$. By symmetry, one can easily find out that there are 6 cases needed to be discussed: $i = j$, $i = i'$, $i = j'$, $j = k$, $j = j'$ and $j = k'$. Nonetheless, it is easy to prove for each case. For example, when $i = j$, then
\begin{align}
    &\sum_{1 \leq i, i', j, j', k, k' \leq n: i = j}\E\Bigg\{\Bigg(\int_0^{\widetilde T_i} \xi_{h, j}\{t, \overline W_i(t), \overline X_i(t)\}\\
    &~~~~\left[d\xi_{q, k}\{t, \overline W_i(t), \overline X_i(t)\} - \xi_{q, k}\{t, \overline W_i(t), \overline X_i(t)\}dN_{C, i}(t) \right]\Bigg)\\
    &~~~\Bigg(\int_0^{\widetilde T_{i'}} \xi_{h, j'}\{t, \overline W_{i'}(t), \overline X_{i'}(t)\}\\
    &~~~~~\left[d\xi_{q, k'}\{t, \overline W_{i'}(t), \overline X_{i'}(t)\} - \xi_{q, k'}\{t, \overline W_{i'}(t), \overline X_{i'}(t)\}dN_{C, i'}(t) \right]\Bigg)\Bigg\}\\
    &=\sum_{1 \leq i, i', j', k, k' \leq n}\E\Bigg\{\Bigg(\int_0^{\widetilde T_i} \xi_{h, i}\{t, \overline W_i(t), \overline X_i(t)\}\\
    &~~~~~\left[d\xi_{q, k}\{t, \overline W_i(t), \overline X_i(t)\} - \xi_{q, k}\{t, \overline W_i(t), \overline X_i(t)\}dN_{C, i}(t) \right]\Bigg)\\
    &~~~\Bigg(\int_0^{\widetilde T_{i'}} \xi_{h, j'}\{t, \overline W_{i'}(t), \overline X_{i'}(t)\}\\
    &~~~~\left[d\xi_{q, k'}\{t, \overline W_{i'}(t), \overline X_{i'}(t)\} - \xi_{q, k'}\{t, \overline W_{i'}(t), \overline X_{i'}(t)\}dN_{C, i'}(t) \right]\Bigg)\Bigg\}\\
    &=\left(\sum_{1 \leq i, i', j', k, k' \leq n: i, i', j', k, k'\text{not equal to each other}} + \sum_{1 \leq i, i', j', k, k' \leq n: i, i', j', k, k' \text{at least two are equal}}\right)\\
    &\E\Bigg\{\left(\int_0^{\widetilde T_i} \xi_{h, i}\{t, \overline W_i(t), \overline X_i(t)\}\left[d\xi_{q, k}\{t, \overline W_i(t), \overline X_i(t)\} - \xi_{q, k}\{t, \overline W_i(t), \overline X_i(t)\}dN_{C, i}(t) \right]\right)\\
    &~~~\Bigg(\int_0^{\widetilde T_{i'}} \xi_{h, j'}\{t, \overline W_{i'}(t), \overline X_{i'}(t)\}\\
    &~~~~~~\left[d\xi_{q, k'}\{t, \overline W_{i'}(t), \overline X_{i'}(t)\} - \xi_{q, k'}\{t, \overline W_{i'}(t), \overline X_{i'}(t)\}dN_{C, i'}(t) \right]\Bigg)\Bigg\}.
\end{align}
For the sum $\sum_{1 \leq i, i', j', k, k' \leq n: i, i', j', k, k'\text{not equal to each other}}$, since $k$ is different from $i, i', j', k'$, by independence and fact that 
\begin{equation}
    \E\left[d\xi_{q, k}\{t, \overline W_i(t), \overline X_i(t)\} - \xi_{q, k}\{t, \overline W_i(t), \overline X_i(t)\}dN_{C, i}(t) \right] = 0,
\end{equation}
it thus follows that $\E\left(\sum_{1 \leq i, i', j', k, k' \leq n: i, i', j', k, k'\text{not equal to each other}}\cdots\right) = 0$. The rest are similar and therefore $\sqrt{n}\cT_1 = o_p(1)$.

We now work on $\sqrt{n}\cT_2$. By linearity of $\Xi$,
\begin{align}
    \cT_2 &= \frac{1}{\sqrt{n}}\sum_{i = 1}^{n}\Xi_i\{\hat H(t; D, \theta), Q_*(t)\} - \frac{1}{n}\sum_{i = 1}^{n}\Xi_i\{H_*(t; D, \theta), Q_*(t)\}\\
    &= \frac{1}{\sqrt{n}}\sum_{i = 1}^{n}\Xi_i\left\{\frac{1}{n}\sum_{j = 1}^n \xi_{h, j}(t), Q_*(t)\right\} \\
    &= \frac{1}{n}\sum_{j = 1}^{n}\Xi_j\left\{\frac{1}{\sqrt{n}}\sum_{i = 1}^n \xi_{h, i}(t), Q_*(t)\right\} \\
    &= \frac{1}{n}\sum_{j = 1}^{n}\left(\Xi_j\left\{\frac{1}{\sqrt{n}}\sum_{i = 1}^n \xi_{h, i}(t), Q_*(t)\right\} - \E\left[\Xi\left\{\frac{1}{\sqrt{n}}\sum_{i = 1}^n \xi_{h, i}(t), Q_*(t)\right\}\Bigg|\cO_i\right]\right) \\
    &~~~+ \frac{1}{\sqrt{n}}\sum_{i = 1}^n \E\left[\Xi\left\{\xi_{h, i}(t), Q_*(t)\right\}\Big|\cO_i\right].
\end{align}
The first term can be shown to be $o_p(1)$ by Chebyshev's inequality following even easier argument than that of $\cT_1$. The second term contributes to the normal distribution if $q_* \not= q_0$ and is equal to zero when $q_* = q_0$. It follows that $\sqrt{n}\cT_3 = o_p(1)$ similarly. $\cT_4$ by \citet{van1997weak} is $o_p\{1 + \sqrt{n}(\theta - \theta_0)\}$. $\cT_5$ together with the first order terms from either $\cT_2$ or $\cT_3$ converges to normal distribution by central limit theorem. $\cT_6 = 0$ by Theorem \ref{thm:driden} when either $h_* = h_0$ or $q_* = q_0$.

\noindent
\textsc{Proof of Part} (b):
We omit since the proof is similar to that of part (a).

\section{Compatibility}\label{sec:compatibility}
We first confirm the model compatibility of the data generating mechanism (DGM) described in Section \ref{sec:simu}, that is, we show that Assumptions \ref{assump:latentcar} - \ref{assump:censoruntestcomplete} hold under this DGM. With the time-invariant covariates, it is easy to see that Assumption \ref{assump:latentcar} and \ref{eq:proximalpos} are satisfied. Assumptions \ref{assump:eventbridge} and \ref{assump:censoringbridge} follow from Section \ref{sec:compatible} since our DGM is simpler than the one considered therein. Assumptions \ref{assump:eventuntestcomplete} holds because both $W$ and $U$ are continuous, of the same dimension and share the same support. 

Next we prove that the concrete first-step estimators for $h$ and $q$ considered in Section \ref{sec:concrete} satisfy Assumptions \ref{assump:uniformcons1} and \ref{assump:if}. It is important to realize that our estimators $\hat H(t)$ and $\hat Q(t)$ are piece-wise constant functions in time $t$, with possible jumps at event times. More importantly, each jump is linear in samples. Therefore, by a similar argument adopted in \citet{martinussen2017instrumental, ying2022structural, ying2022structural2}, one can show that both $\hat H(t)$ and $\hat Q(t)$ are not only asymptotically linear, but indeed linear in samples. There is indeed no remainder term after linear expansion and therefore the remainder total variation conditions in Assumption \ref{assump:if} hold automatically. Moreover, our estimators considered are of bounded variation and thus in a Donsker class. By applying the Donsker Theorem, the supremum converge conditions and rate conditions in Assumptions \ref{assump:uniformcons1} and \ref{assump:if} hold.

\end{document}

%% file: notation.tex
\newtheorem{thm}{Theorem}

\newtheorem{assump}{Assumption}

\theoremstyle{remark}

\newtheorem{rem}{Remark}

\def\beq{\begin{equation}} 
\def\eeq{\end{equation}}
\def\beqn{\begin{eqnarray*}}
\def\eeqn{\end{eqnarray*}}
\def\Bitem{\begin{itemize}\setlength{\itemsep}{.2in}}
\def\bitem{\begin{itemize}\setlength{\itemsep}{.05in}}
\def\eitem{\end{itemize}}
\def\Benum{\begin{enumerate}\setlength{\itemsep}{.2in}}
\def\benum{\begin{enumerate}\setlength{\itemsep}{.05in}}
\def\eenum{\end{enumerate}}
\def\bmult{\begin{multline*}}
\def\emult{\end{multline*}}
\def\bcenter{\begin{center}}
\def\ecenter{\end{center}}
\def\bframe{\begin{frame}}
\def\eframe{\end{frame}}

\def\cN{\mathcal{N}}
\def\cO{\mathcal{O}}
\def\cP{\mathcal{P}}

\def\cT{\mathcal{T}}

\def\bbM{\mathbb{M}}

\newcommand{\E}{\operatorname{\mathbb{E}}}
\renewcommand{\P}{\operatorname{\mathbb{P}}}
\newcommand{\Var}{\operatorname{Var}}

\def\eps{\varepsilon}

\def\1{\mathbbm{1}}

%% file: lpci_ref.bib
@article{ai2003efficient,
  title={Efficient estimation of models with conditional moment restrictions containing unknown functions},
  author={Ai, Chunrong and Chen, Xiaohong},
  journal={Econometrica},
  volume={71},
  number={6},
  pages={1795--1843},
  year={2003},
  publisher={Wiley Online Library}
}

@article{andrews2017examples,
	Author = {Andrews, Donald WK},
	Journal = {Journal of Econometrics},
	Pages = {213--220},
	Publisher = {Elsevier},
	Title = {Examples of $\uppercase{L}^2$-complete and boundedly-complete distributions},
	Volume = {199},
	Year = {2017}
}

@article{chen2014local,
  title={Local identification of nonparametric and semiparametric models},
  author={Chen, Xiaohong and Chernozhukov, Victor and Lee, Sokbae and Newey, Whitney K},
  journal={Econometrica},
  volume={82},
  number={2},
  pages={785--809},
  year={2014},
  publisher={Wiley Online Library}
}

@article{cui2020semiparametric,
  title={Semiparametric proximal causal inference},
  author={Cui, Yifan and Pu, Hongming and Shi, Xu and Miao, Wang and Tchetgen Tchetgen, Eric J.},
  journal={arXiv preprint arXiv:2011.08411},
  year={2020}
}

@article{d2011completeness,
  title={On the completeness condition in nonparametric instrumental problems},
  author={D'Haultfoeuille, Xavier},
  journal={Econometric Theory},
  pages={460--471},
  year={2011},
  publisher={JSTOR}
}

@article{hu2018nonparametric,
  title={Nonparametric identification using instrumental variables: sufficient conditions for completeness},
  author={Hu, Yingyao and Shiu, Ji-Liang},
  journal={Econometric Theory},
  volume={34},
  number={3},
  pages={659--693},
  year={2018},
  publisher={Cambridge University Press}
}

@article{imbens2021controlling,
  title={Controlling for Unmeasured Confounding in Panel Data Using Minimal Bridge Functions: From Two-Way Fixed Effects to Factor Models},
  author={Imbens, Guido and Kallus, Nathan and Mao, Xiaojie},
  journal={arXiv preprint arXiv:2108.03849},
  year={2021}
}

@article{kallus2021causal,
  title={Causal Inference Under Unmeasured Confounding With Negative Controls: A Minimax Learning Approach},
  author={Kallus, Nathan and Mao, Xiaojie and Uehara, Masatoshi},
  journal={arXiv preprint arXiv:2103.14029},
  year={2021}
}

@inproceedings{mastouri2021proximal,
  title={Proximal causal learning with kernels: Two-stage estimation and moment restriction},
  author={Mastouri, Afsaneh and Zhu, Yuchen and Gultchin, Limor and Korba, Anna and Silva, Ricardo and Kusner, Matt and Gretton, Arthur and Muandet, Krikamol},
  booktitle={International Conference on Machine Learning},
  pages={7512--7523},
  year={2021},
  organization={PMLR}
}

@article{newey2003instrumental,
  title={Instrumental variable estimation of nonparametric models},
  author={Newey, Whitney K and Powell, James L},
  journal={Econometrica},
  volume={71},
  number={5},
  pages={1565--1578},
  year={2003},
  publisher={Wiley Online Library}
}

@article{shi2020selective,
	author = {Shi, Xu and Miao, Wang and Tchetgen Tchetgen, Eric J.},
	da = {2020/12/01},
	doi = {10.1007/s40471-020-00243-4},
	id = {Shi2020},
	isbn = {2196-2995},
	journal = {Current Epidemiology Reports},
	number = {4},
	pages = {190--202},
	title = {A Selective Review of Negative Control Methods in Epidemiology},
	ty = {JOUR},
	url = {https://doi.org/10.1007/s40471-020-00243-4},
	volume = {7},
	year = {2020}}

@book{van1997weak,
  title={Weak Convergence and Empirical Processes},
  author={Van der Vaart, Aad W and Wellner, Jon A},
  volume={3},
  year={1997},
  publisher={Springer New York}
}


%% file: ref.bib
@incollection{aalen1980model,
  title={A model for nonparametric regression analysis of counting processes},
  author={Aalen, Odd},
  booktitle={Mathematical statistics and probability theory},
  pages={1--25},
  year={1980},
  publisher={Springer}
}

@article{aalen1989linear,
  title={A linear regression model for the analysis of life times},
  author={Aalen, Odd O},
  journal={Statistics in Medicine},
  volume={8},
  number={8},
  pages={907--925},
  year={1989},
  publisher={Wiley Online Library}
}

@book{andersen2012statistical,
  title={Statistical Models Based on Counting Processes},
  author={Andersen, Per K and Borgan, Ornulf and Gill, Richard D and Keiding, Niels},
  year={2012},
  publisher={Springer Science \& Business Media}
}

@article{austin2020review,
  title={A review of the use of time-varying covariates in the Fine-Gray subdistribution hazard competing risk regression model},
  author={Austin, Peter C and Latouche, Aur{\'e}lien and Fine, Jason P},
  journal={Statistics in Medicine},
  volume={39},
  number={2},
  pages={103--113},
  year={2020},
  publisher={Wiley Online Library}
}

@article{beyersmann2008time,
  title={Time-dependent covariates in the proportional subdistribution hazards model for competing risks},
  author={Beyersmann, Jan and Schumacher, Martin},
  journal={Biostatistics},
  volume={9},
  number={4},
  pages={765--776},
  year={2008},
  publisher={Oxford University Press}
}

@article{byar1980choice,
  title={The choice of treatment for cancer patients based on covariate information.},
  author={Byar, David P and Green, Sylvan B},
  journal={Bulletin du cancer},
  volume={67},
  number={4},
  pages={477--490},
  year={1980}
}

@article{carrasco2007linear,
  title={Linear inverse problems in structural econometrics estimation based on spectral decomposition and regularization},
  author={Carrasco, Marine and Florens, Jean-Pierre and Renault, Eric},
  journal={Handbook of Econometrics},
  volume={6},
  pages={5633--5751},
  year={2007},
  publisher={Elsevier}
}

@article{cox1972regression,
  title={Regression models and life-tables},
  author={Cox, David R},
  journal={Journal of the Royal Statistical Society: Series B (Methodological)},
  volume={34},
  number={2},
  pages={187--202},
  year={1972},
  publisher={Wiley Online Library}
}

@article{cox1975partial,
  title={Partial likelihood},
  author={Cox, David R},
  journal={Biometrika},
  volume={62},
  number={2},
  pages={269--276},
  year={1975},
  publisher={Oxford University Press}
}

@article{darolles2011nonparametric,
  title={Nonparametric instrumental regression},
  author={Darolles, Serge and Fan, Yanqin and Florens, Jean-Pierre and Renault, Eric},
  journal={Econometrica},
  volume={79},
  number={5},
  pages={1541--1565},
  year={2011},
  publisher={Wiley Online Library}
}

@article{deaner2018proxy,
  title={Proxy controls and panel data},
  author={Deaner, Ben},
  journal={arXiv preprint arXiv:1810.00283},
  year={2018}
}

@article{dukes2021proximal,
  title={Proximal mediation analysis},
  author={Dukes, Oliver and Shpitser, Ilya and Tchetgen Tchetgen, Eric J.},
  journal={arXiv preprint arXiv:2109.11904},
  year={2021}
}

@book{efron1994introduction,
  title={An Introduction to the Bootstrap},
  author={Efron, Bradley and Tibshirani, Robert J},
  year={1994},
  publisher={CRC press}
}

@article{fine1999proportional,
  title={A proportional hazards model for the subdistribution of a competing risk},
  author={Fine, Jason P and Gray, Robert J},
  journal={Journal of the American Statistical Association},
  volume={94},
  number={446},
  pages={496--509},
  year={1999},
  publisher={Taylor \& Francis}
}

@article{gill1990survey,
  title={A survey of product-integration with a view toward application in survival analysis},
  author={Gill, Richard D and Johansen, Soren},
  journal={The Annals of Statistics},
  volume={18},
  number={4},
  pages={1501--1555},
  year={1990},
  publisher={Institute of Mathematical Statistics}
}

@inproceedings{ghassami2022minimax,
  title={Minimax Kernel Machine Learning for a Class of Doubly Robust Functionals with Application to Proximal Causal Inference},
  author={Ghassami, AmirEmad and Ying, Andrew and Shpitser, Ilya and Tchetgen Tchetgen, Eric J.},
  booktitle={International Conference on Artificial Intelligence and Statistics},
  pages={7210--7239},
  year={2022},
  organization={PMLR}
}

@article{heitjan1991ignorability,
  title={Ignorability and coarse data},
  author={Heitjan, Daniel F and Rubin, Donald B},
  journal={The Annals of Statistics},
  pages={2244--2253},
  volume={19},
  number={4},
  year={1991},
  publisher={JSTOR}
}

@article{hou2021treatment,
  title={Treatment effect estimation under additive hazards models with high-dimensional confounding},
  author={Hou, Jue and Bradic, Jelena and Xu, Ronghui},
  journal={Journal of the American Statistical Association},
  volume={118},
  number={541},
  pages={327--342},
  year={2023},
  publisher={Taylor \& Francis}
}

@book{kalbfleisch2011statistical,
  title={The Statistical Analysis of Failure Time Data},
  author={Kalbfleisch, John D and Prentice, Ross L},
  year={2011},
  publisher={John Wiley \& Sons}
}

@article{kaplan1958nonparametric,
  title={Nonparametric estimation from incomplete observations},
  author={Kaplan, Edward L and Meier, Paul},
  journal={Journal of the American Statistical Association},
  volume={53},
  number={282},
  pages={457--481},
  year={1958},
  publisher={Taylor \& Francis}
}

@book{kosorok2008introduction,
  title={Introduction to Empirical Processes and Semiparametric Inference},
  author={Kosorok, Michael R},
  volume={61},
  year={2008},
  publisher={Springer}
}

@book{kress1989linear,
  title={Linear Integral Equations},
  author={Kress, Rainer and Maz'ya, V and Kozlov, V},
  volume={82},
  year={1989},
  publisher={Springer}
}

@article{lo1991bayesian,
  title={Bayesian bootstrap clones and a biometry function},
  author={Lo, Albert Y},
  journal={Sankhy{\=a}: The Indian Journal of Statistics, Series A},
  pages={320--333},
  volume={53},
  number={3},
  year={1991},
  publisher={JSTOR}
}

@book{martinussen2006dynamic,
  title={Dynamic Regression Models for Survival Data},
  author={Martinussen, Torben and Scheike, Thomas H},
  volume={1},
  year={2006},
  publisher={Springer}
}

@article{martinussen2017instrumental,
  title={Instrumental variables estimation of exposure effects on a time-to-event endpoint using structural cumulative survival models},
  author={Martinussen, Torben and Vansteelandt, Stijn and Tchetgen Tchetgen, Eric J. and Zucker, David M},
  journal={Biometrics},
  volume={73},
  number={4},
  pages={1140--1149},
  year={2017},
  publisher={Wiley Online Library}
}

@article{miao2018identifying,
  title={Identifying Causal Effects with Proxy Variables of an Unmeasured Confounder},
  author={Miao, Wang and Geng, Zhi and Tchetgen Tchetgen, Eric J},
  journal={Biometrika},
  volume={105},
  number={4},
  pages={987--993},
  year={2018},
  publisher={Oxford University Press}
}

@article{p2012honolulu,
  title={The Honolulu-Asia Aging Study: epidemiologic and neuropathologic research on cognitive impairment},
  author={P Gelber, Rebecca and J Launer, Lenore and R White, Lon},
  journal={Current Alzheimer Research},
  volume={9},
  number={6},
  pages={664--672},
  year={2012},
  publisher={Bentham Science Publishers}
}

@article{qi2021proximal,
  title={Proximal learning for individualized treatment regimes under unmeasured confounding},
  author={Qi, Zhengling and Miao, Rui and Zhang, Xiaoke},
  journal={arXiv preprint arXiv:2105.01187},
  year={2021}
}

@article{rotnitzky2005inverse,
  title={Inverse probability weighted estimation in survival analysis},
  author={Rotnitzky, Andrea and Robins, James},
  journal={Encyclopedia of Biostatistics},
  volume={4},
  pages={2619--2625},
  year={2005},
  publisher={Citeseer}
}

@article{royston2013restricted,
  title={Restricted mean survival time: an alternative to the hazard ratio for the design and analysis of randomized trials with a time-to-event outcome},
  author={Royston, Patrick and Parmar, Mahesh KB},
  journal={BMC Medical Research Methodology},
  volume={13},
  number={1},
  pages={1--15},
  year={2013},
  publisher={BioMed Central}
}

@article{rubin1981bayesian,
  title={The Bayesian Bootstrap},
  author={Rubin, Donald B},
  journal={The Annals of Statistics},
  volume={9},
  number={1},
  pages={130--134},
  year={1981},
  publisher={Institute of Mathematical Statistics}
}

@book{shao2012jackknife,
  title={The Jackknife and Bootstrap},
  author={Shao, Jun and Tu, Dongsheng},
  year={2012},
  publisher={Springer Science \& Business Media}
}

@article{singh2020kernel,
  title={Kernel methods for unobserved confounding: Negative controls, proxies, and instruments},
  author={Singh, Rahul},
  journal={arXiv preprint arXiv:2012.10315},
  year={2020}
}

@article{tchetgen2020introduction,
  title={An introduction to proximal causal learning},
  author={Tchetgen Tchetgen, Eric J. and Ying, Andrew and Cui, Yifan and Shi, Xu and Miao, Wang},
  journal={arXiv preprint arXiv:2009.10982},
  year={2020}
}

@book{tsiatis2006semiparametric,
  title={Semiparametric Theory and Missing Data},
  author={Tsiatis, Anastasios A},
  year={2006},
  publisher={Springer}
}

@book{van2003unified,
  title={Unified Methods for Censored Longitudinal Data and Causality},
  author={Van der Laan, Mark J and Laan, MJ and Robins, James M},
  year={2003},
  publisher={Springer Science \& Business Media}
}

@article{wang2022doubly,
  title={Doubly Robust Estimation under Covariate-induced Dependent Left Truncation},
  author={Wang, Yuyao and Ying, Andrew and Xu, Ronghui},
  journal={arXiv preprint arXiv:2208.06836},
  year={2022}
}

@article{ying2019two,
  title={Two-stage residual inclusion for survival data and competing risks—An instrumental variable approach with application to SEER-Medicare linked data},
  author={Ying, Andrew and Xu, Ronghui and Murphy, James},
  journal={Statistics in Medicine},
  volume={38},
  number={10},
  pages={1775--1801},
  year={2019},
  publisher={Wiley Online Library}
}

@article{ying2021proximal,
  title={Proximal causal inference for complex longitudinal studies},
  author={Ying, Andrew and Miao, Wang and Shi, Xu and Tchetgen Tchetgen, Eric J},
  journal={Journal of the Royal Statistical Society Series B: Statistical Methodology},
  volume={85},
  number={3},
  pages={684--704},
  year={2023},
  publisher={Oxford University Press US}
}

@article{ying2022proximal,
  title={Proximal Causal Inference for Marginal Counterfactual Survival Curves},
  author={Ying, Andrew and Cui, Yifan and Tchetgen Tchetgen, Eric J.},
  journal={arXiv preprint arXiv:2204.13144},
  year={2022}
}

@article{ying2022structural,
  title={Structural cumulative survival models for estimation of treatment effects accounting for treatment switching in randomized experiments},
  author={Ying, Andrew and Tchetgen Tchetgen, Eric J},
  journal={Biometrics},
  volume={79},
  number={3},
  pages={1597--1609},
  year={2023},
  publisher={Wiley Online Library}
}

@article{ying2022structural2,
  title={Structural Cumulative Survival Models for Robust Estimation of Treatment Effects Accounting for Treatment Switching in Randomized Experiments},
  author={Ying, Andrew},
  journal={arXiv preprint arXiv:2204.13219},
  year={2022}
}

@article{ying2023cautionary,
  title={A Cautionary Note on Doubly Robust Estimators Involving Continuous-time Structure},
  author={Ying, Andrew},
  journal={arXiv preprint arXiv:2302.06739},
  year={2023}
}
